\documentclass[11pt]{article}

\usepackage[margin=1in]{geometry}
\usepackage{amsmath,amssymb,bm}
\usepackage{booktabs}
\usepackage{tabularx}
\usepackage{graphicx}
\usepackage{xcolor}
\usepackage{hyperref}
\usepackage{microtype}
\usepackage{siunitx}
\usepackage{placeins}
\usepackage{multirow}
\usepackage{authblk}
\usepackage{orcidlink}
\usepackage{fancyhdr}
\usepackage{tikz}
\usepackage{adjustbox}
\usetikzlibrary{arrows.meta,positioning,fit,calc}

\hypersetup{colorlinks=true,linkcolor=blue,citecolor=blue,urlcolor=blue}

\usepackage[nameinlink,capitalise]{cleveref}
\usepackage[labelfont={bf,it},textfont=it]{caption}

\newcommand{\bu}{\bm{u}}

\newcommand{\bmpar}{\bm{m}}
\newcommand{\bq}{\bm{q}}
\newcommand{\Rop}{\mathcal{R}}
\newcommand{\Fop}{\mathcal{F}}
\newcommand{\FNOp}{\mathcal{F}_{\mathrm{NOp}}}
\newcommand{\EQ}{E_Q}
\newcommand{\Ehat}{\widehat E_Q}

\title{Adapting neural operators for mechanics decisions under changing operating conditions}
\author[1]{Prashant K. Jha\thanks{\raggedright Corresponding author. \\ Emails: PKJ, \href{mailto:pjha.sci@gmail.com}{pjha.sci@gmail.com}; KE, \href{mailto:koffi@math.ucla.edu}{koffi@math.ucla.edu}; IG, \href{mailto:ian.galloway@mines.sdsmt.edu}{ian.galloway@mines.sdsmt.edu}; HA, \href{mailto:henry.anderson@mines.sdsmt.edu}{henry.anderson@mines.sdsmt.edu}.}\,\orcidlink{0000-0003-2158-364X}}
\author[2]{Koffi Enakoutsa\,\orcidlink{0009-0008-7859-2030}}
\author[1]{Ian Galloway\,\orcidlink{0009-0002-6428-8624}}
\author[1]{Henry Anderson\,\orcidlink{0009-0009-7685-9878}}
\affil[1]{Department of Mechanical Engineering, South Dakota School of Mines and Technology, Rapid City, SD 57701, USA}
\affil[2]{Department of Mathematics, University of California, Los Angeles, Los Angeles, CA 90095, USA}
\date{}

\begin{document}
\maketitle

\begin{abstract}
Neural operators can accelerate repeated nonlinear mechanics calculations, but their accuracy can deteriorate as operating conditions move beyond the training range. This work studies whether high-fidelity solutions acquired during use can be reused to adapt a neural operator and improve subsequent mechanics-based command selection. Two hard-magnetic soft-material systems are simulated using high-fidelity finite-element (FE) models, providing reference solutions for evaluating surrogate predictions and selected commands. A neural operator predicts deformation from known material, loading, and magnetic-field inputs, while an empirical error estimator determines which predictions may be used for command selection. Selected FE evaluations supplement these predictions, and their complete loading paths are retained for periodic updates of the neural operator and estimator. In both examples, the fixed operator loses substantial accuracy when stiffness and loading move outside the training range. Updates using 16 acquired paths recover much of the lost accuracy while preserving accuracy in the nominal regime. Under the same FE evaluation budget, the updated operators also improve command selection, although the benefit varies with the operating condition. Error estimation is less consistent, with inaccurate predictions sometimes accepted and accurate predictions rejected. These results demonstrate that reusing high-fidelity loading paths can extend the useful operating range of a neural operator. However, improved forward accuracy alone does not guarantee reliable prediction acceptance, highlighting prediction-specific error assessment as a separate requirement for trustworthy decision making.
\end{abstract}


\medskip


\section{Introduction}
\label{sec:introduction}

A computational twin used for mechanics-based decision making must repeatedly evaluate the response of a physical system as operating conditions and requested actions change. High-fidelity finite-element (FE) models can provide the displacement, strain, and other mechanical quantities needed for these decisions from known boundary conditions and material properties, but repeated FE calculations may be too expensive for practical use. This difficulty is pronounced for field-responsive soft materials, where finite-strain equilibrium calculations commonly require incremental loading and iterative solution of a nonlinear problem at each increment. Reduced-order and learned models can reduce this computational cost by approximating the forward response in place of repeated FE solutions \cite{rasheed2020digital,kapteyn2021digital,torzoni2024civil}.

For forward models governed by parameterized partial differential equations (PDEs), neural operators learn the map from model inputs to solution fields. Representative architectures include DeepONet \cite{lu2021deeponet}, the Fourier neural operator \cite{li2021fno}, and projection-based approaches that represent the solution in a low-dimensional basis and learn the corresponding coefficients \cite{BhattacharyaHosseiniKovachkiEtAl2020}. Once trained, these operators can approximate the solution at new inputs without repeatedly solving the governing PDEs. This capability has motivated their use in forward prediction, inverse problems, and design, where many evaluations of the same solution map are required \cite{jha2026mathematics,kobayashi2024digital,cao2023residual,jha_koffi_bookchapter2026}.

The usefulness of this approximation depends on whether its predictions are sufficiently accurate for the quantities of interest (QoIs) used in a decision. Neural-operator error depends on the solution representation, network approximation, and sampling, as described by error analyses for DeepONet and Fourier neural operators \cite{lanthaler2022deeponeterror,kovachki2021fnoerror}. Accuracy is usually assessed using test samples drawn from the distribution represented during training. Operating conditions encountered during use may move beyond this nominal regime, however, and neural-operator accuracy can deteriorate under the resulting distribution shift \cite{takamoto2022pdebench,gupta2023pdearena,benitez2023oodrisk}. Moreover, aggregate test accuracy does not determine whether an individual prediction satisfies the accuracy requirements of a particular QoI or decision \cite{jha_koffi_bookchapter2026}. In nonlinear mechanics, an inaccurate displacement prediction can misrepresent the achieved translation or rotation, incorrectly indicate that a strain constraint is satisfied, and change the action selected from a discrete set. Evaluating every candidate action by FE avoids relying on these surrogate predictions but removes much of their computational benefit.

Prediction-specific error assessment and refinement address this limitation in several ways. Goal-oriented estimation evaluates approximation error through a selected output rather than only through a global field norm \cite{prudhomme1999goal, oden2001goal, oden2002estimation, becker2001optimal,jha2022goal}. Residual-based methods use additional variational calculations to improve individual neural-operator predictions \cite{cao2023residual,jha2024residual}, while a posteriori approaches seek computable estimates of their errors \cite{fanaskov2024functionalaposteriori,qiu2026variational}. Conformal operator-learning methods provide calibrated uncertainty under their sampling assumptions, although this does not guarantee accuracy conditional on every shifted input \cite{ma2024conformal,barber2021limits}. These methods provide information for assessing or refining a prediction, but repeated use under changing conditions also raises the question of how to improve the surrogate itself.

High-fidelity FE calculations obtained during use can provide data for this improvement. Adaptive operator learning introduces additional high-fidelity data in regions encountered during an inverse problem \cite{gao2024adaptive}, while hierarchical and multifidelity methods combine approximations of different cost and fidelity \cite{haasdonk2023certified,lu2022multifidelitydno,howard2023multifidelitydeeponet}. For repeated mechanics decisions, the value of additional data must be assessed beyond the resulting reduction in forward error. It is also necessary to determine whether the updated predictions improve action selection and whether their errors can still be estimated reliably. This motivates studying the reuse of high-fidelity solutions together with prediction-specific error assessment and decision quality under changing known operating conditions.

The present study addresses four related questions. First, does a neural operator that is accurate on held-out nominal inputs remain accurate when known stiffness and mechanical loads move beyond its training range? Second, can a small number of FE loading paths collected under these shifted conditions recover the lost forward accuracy without degrading accuracy in the nominal regime? Third, does improved forward accuracy lead to better action selection under the same FE budget? Fourth, does adaptation of the forward operator also improve the reliability of its error estimator?

These questions are examined for two hard-magnetic soft-material (hMSM) structures. Hard-magnetic particles embedded in a compliant matrix retain a prescribed remanent magnetization, allowing an applied field to deform the structure without direct mechanical contact \cite{zhao2019mechanics,kim2022magnetic,kim2018printing}. The deformation produced by a given field also depends on material stiffness and mechanical loading, so a field selected under one condition need not produce the requested response under another. The examples draw on prior work in hMSM topology and joint material--structural optimization \cite{zhao2022topology,galloway_jha_hmsm2026} and the MatTO framework \cite{galloway2026mattoarticle}. Neural-network surrogates have also been used to search hMSM morphology and magnetization for prescribed magnetic fields \cite{karacakol2025datadriven}. The physical systems are simulated using high-fidelity FE models, with material properties and mechanical loads supplied as known inputs. This setting permits direct assessment of surrogate errors and selected actions without experimental uncertainty, sensor noise, or hidden-parameter inference. The tests vary known stiffness and mechanical loading beyond the training range while keeping the magnetic-command range unchanged.

\begin{figure}[h]
\centering
\includegraphics[width=6.41in]{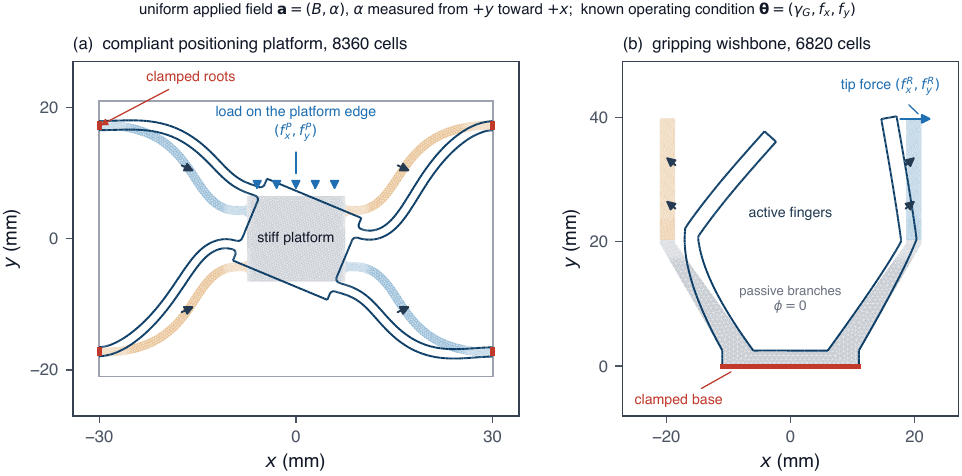}
\vspace{-10pt}
\caption{The two hard-magnetic soft-material structures considered in this study. (a) A compliant positioning platform supported by four magnetically active ligaments and clamped at the outer roots. (b) A gripping wishbone with two magnetically active fingers supported by passive lower branches and clamped along the base. Arrows within the active regions show remanent magnetization, and grey denotes passive material. Dark outlines show representative FE deformations; the platform displacement is magnified twofold, while the wishbone is shown at true scale.}
\label{fig:work}
\end{figure}

The two structures in \cref{fig:work} provide different command-selection tasks. The compliant positioning platform is supported by four curved hMSM ligaments. Its horizontal translation and rotation define the requested pose, while vertical drift and a regional strain measure determine admissibility. The gripping wishbone is based on the gripper geometry studied in \cite{galloway_jha_hmsm2026}. Its two magnetically active fingers are supported by passive branches, and its response here is characterized by aperture, lateral position, vertical mismatch, and root strain. In both examples, a magnetic command specifies the magnitude and direction of a uniform applied field. The objective is to select a command that produces the requested response while satisfying the prescribed mechanical admissibility criteria.

A projection-based neural operator is constructed separately for each structure. The displacement is represented in a fixed proper orthogonal decomposition (POD) basis, and a neural network learns the dependence of its coefficients on material, mechanical-loading, and magnetic-field inputs. For each operating condition, the operator predicts the displacement under every candidate magnetic command. The QoIs extracted from these predictions are used to evaluate command performance and admissibility. An empirical estimator uses the nonlinear residual and one linearized residual solve to estimate the QoI error following \cite{jha2022goal, oden2002estimation} and determine which predictions may be used. The linearized solve supplies information for error estimation; it does not correct the predicted displacement.

The accepted surrogate predictions are supplemented by four FE evaluations at each operating condition. The commands evaluated by FE are selected either using the error estimator or by random sampling, with the same four-path budget for both strategies. Their FE QoIs enter the current command calculation, while the complete sequence of equilibrium solutions along each loading path is retained for adaptation. The neural operator is updated after 8 and 16 acquired paths, and the estimator is refitted as the operator and available FE data change. Fixed and updated operators are then compared on held-out operating conditions to assess forward accuracy, estimator reliability, and command selection. The platform and wishbone use separate POD bases, training data, operators, and estimators.

The numerical studies show that accuracy within the training regime does not persist under the prescribed changes in stiffness and loading. In both examples, periodic updates using 16 acquired FE paths recover much of the lost forward accuracy while preserving accuracy in the nominal regime. These improvements also lead to better command selection under the same FE budget, although the benefit varies with the operating condition. Updating the forward operator and refitting its estimator do not consistently produce reliable acceptance decisions: inaccurate predictions may still be accepted, and accurate predictions may be rejected. The comparisons also do not establish an advantage of estimator-guided acquisition over random acquisition.

Together, these findings support reusing FE loading paths to maintain the usefulness of a neural operator under changing operating conditions. The benefit extends beyond reducing forward error to improving subsequent mechanics-based decisions. However, adaptation alone does not establish when the updated predictions can be accepted. The study therefore demonstrates the value of high-fidelity data reuse for forward adaptation and identifies reliable prediction-specific error assessment as a separate requirement for its use in decision making.

The rest of the article is organized as follows. \cref{sec:problem} defines the forward model and formulates the two command-selection problems. \cref{sec:method} presents the neural operator, QoI-error estimator, selective FE evaluation, and periodic adaptation. \cref{sec:platform,sec:wishbone} present the numerical setup and results for the platform and wishbone, respectively. \cref{sec:discussion,sec:conclusion} discuss the results and summarize the conclusions.

\section{Forward mechanics and command-selection problems}
\label{sec:problem}

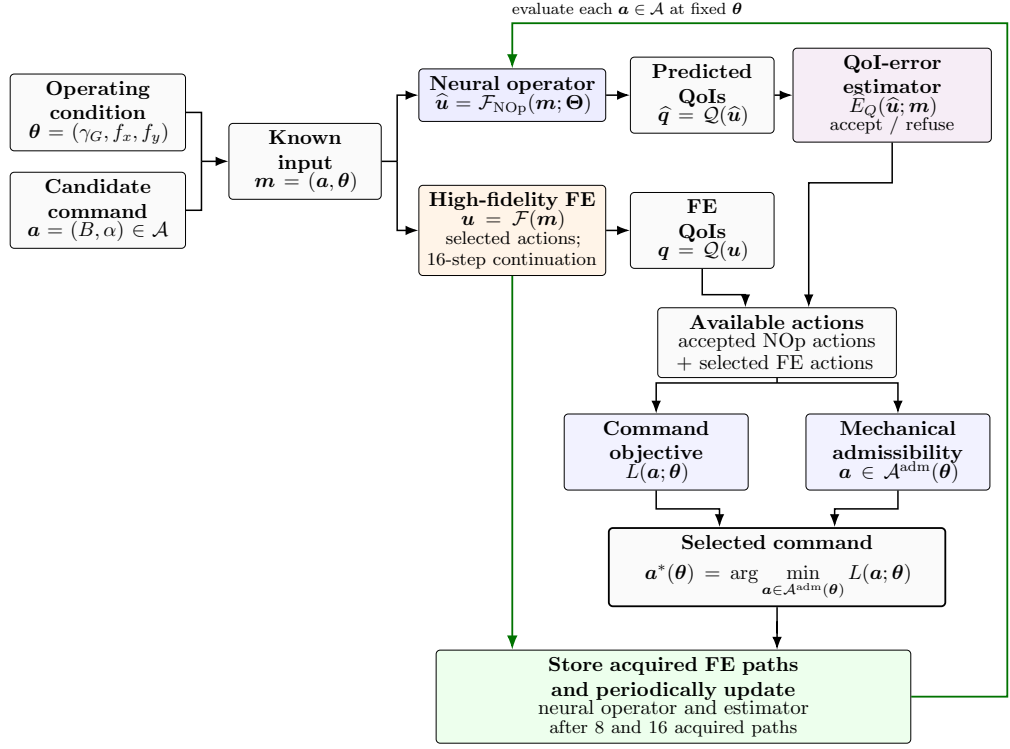
\begin{figure}[h]
\centering
\begin{adjustbox}{max width=0.8\linewidth}
\begin{tikzpicture}[
  x=1cm,
  y=1cm,
  >=Latex,
  font=\small,
  box/.style={
    draw,
    rounded corners=2pt,
    align=center,
    minimum height=0.9cm,
    inner sep=3.5pt,
    fill=gray!4
  },
  inputbox/.style={box,text width=2.8cm},
  processbox/.style={box,text width=2.9cm},
  qbox/.style={box,text width=2.25cm},
  decisionbox/.style={box,text width=2.95cm,fill=blue!5},
  updatebox/.style={box,text width=8.0cm,fill=green!7},
  arrow/.style={->,line width=0.75pt},
  uarrow/.style={->,line width=0.9pt,green!45!black},
  plain/.style={line width=0.75pt}
]

\node[inputbox] (theta) at (0,0.9) {
  \textbf{Operating\\condition}\\[-1mm]
  $\bm\theta=(\gamma_G,f_x,f_y)$
};

\node[inputbox] (action) at (0,-0.8) {
  \textbf{Candidate\\command}\\[-1mm]
  $\bm a=(B,\alpha)\in\mathcal A$
};

\coordinate (merge) at (1.8,0.05);

\draw[plain] (theta.east) -- (1.8,0.9) -- (merge);
\draw[plain] (action.east) -- (1.8,-0.8) -- (merge);

\node[processbox,text width=2.4cm] (input) at (3.6,0.05) {
  \textbf{Known\\input}\\[-1mm]
  $\bmpar=(\bm a,\bm\theta)$
};

\draw[arrow] (merge) -- (input.west);

\coordinate (split) at (5.2,0.05);
\draw[plain] (input.east) -- (split);

\node[processbox,fill=blue!7,text width=3.0cm] (nop) at (7.2,1.20) {
  \textbf{Neural operator}\\[-1mm]
  $\widehat{\bu}=\FNOp(\bmpar;\bm\Theta)$
};

\node[processbox,fill=orange!9,text width=3.0cm] (fe) at (7.2,-1.15) {
  \textbf{High-fidelity FE}\\[-1mm]
  $\bu=\mathcal F(\bmpar)$\\[-1mm]
  {\footnotesize selected actions;}\\[-0.5mm]
  {\footnotesize 16-step continuation}
};

\draw[arrow] (split) |- (nop.west);
\draw[arrow] (split) |- (fe.west);

\node[qbox] (qhat) at (10.5,1.20) {
  \textbf{Predicted}\\
  \textbf{QoIs}\\[-1mm]
  $\widehat{\bq}=\mathcal Q(\widehat{\bu})$
};

\node[qbox] (qfe) at (10.5,-1.15) {
  \textbf{FE}\\
  \textbf{QoIs}\\[-1mm]
  $\bq=\mathcal Q(\bu)$
};

\draw[arrow] (nop.east) -- (qhat.west);
\draw[arrow] (fe.east) -- (qfe.west);

\node[decisionbox,fill=violet!7,text width=3.2cm] (est) at (13.8,1.20) {
  \textbf{QoI-error\\estimator}\\[-1mm]
  $\widehat E_Q(\widehat{\bu};\bmpar)$\\[-1mm]
  {\footnotesize accept / refuse}
};

\draw[arrow] (qhat.east) -- (est.west);

\node[processbox,text width=3.9cm] (avail) at (11.8,-3.1) {
  \textbf{Available actions}\\[-1mm]
  accepted NOp actions\\[-0.5mm]
  $+$ selected FE actions
};

\draw[arrow]
  (qfe.south)
  -- ++(0,-0.45)
  -| ([xshift=-0.55cm]avail.north);

\draw[arrow]
  (est.south)
  -- ++(0,-0.70)
  -| ([xshift=0.55cm]avail.north);

\node[decisionbox] (loss) at (9.7,-5.0) {
  \textbf{Command\\objective}\\[-1mm]
  $L(\bm a;\bm\theta)$
};

\node[decisionbox] (adm) at (13.9,-5.0) {
  \textbf{Mechanical\\admissibility}\\[-1mm]
  $\bm a\in\mathcal A^{\mathrm{adm}}(\bm\theta)$
};

\coordinate (decisionSplit) at (11.8,-3.8);
\draw[plain] (avail.south) -- (decisionSplit);
\draw[arrow] (decisionSplit) -- (9.7,-3.8) -- (loss.north);
\draw[arrow] (decisionSplit) -- (13.9,-3.8) -- (adm.north);

\node[processbox,text width=5.6cm,line width=0.9pt] (select) at (11.8,-7) {
  \textbf{Selected command}\\[1mm]
  $\displaystyle
  \bm a^*(\bm\theta)
  =
  \arg\min_{\bm a\in\mathcal A^{\mathrm{adm}}(\bm\theta)}
  L(\bm a;\bm\theta)$
};

\draw[arrow]
  (loss.south)
  -- ++(0,-0.32)
  -| ([xshift=-1.0cm]select.north);

\draw[arrow]
  (adm.south)
  -- ++(0,-0.32)
  -| ([xshift=1.0cm]select.north);

\node[updatebox] (update) at (10.0,-9.25) {
  \textbf{Store acquired FE paths and periodically update}\\[-1mm]
  neural operator and estimator\\[-1mm]
  {\footnotesize after 8 and 16 acquired paths}
};

\draw[arrow]
  (select.south)
  -- ++(0,-0.45)
  -- (11.8,-7.95)
  -- (11.8,-8.35)
  -- ([xshift=1.8cm]update.north);

\draw[uarrow]
  (fe.south)
  -- ([xshift=-2.8cm]update.north);

\coordinate (feedbackR) at (15.8,-9.25);

\draw[uarrow]
  (update.east)
  -- (feedbackR)
  -- (feedbackR |- 7.2,2.45)
  -- (7.2,2.45)
  -- (nop.north);

\node[font=\scriptsize] at (9.2,2.7) {
  evaluate each $\bm a\in\mathcal A$ at fixed $\bm\theta$
};

\end{tikzpicture}
\end{adjustbox}

\caption{
Reference and surrogate-assisted command-selection sequence. For a fixed operating condition $\bm\theta$, each candidate command $\bm a\in\mathcal A$ defines the known forward input $\bmpar=(\bm a,\bm\theta)$. The neural operator predicts the displacement and associated mechanics QoIs, while selected commands are evaluated using the high-fidelity FE model. The QoI-error estimator determines which neural-operator predictions may enter the command calculation. The available set comprises commands with accepted neural-operator predictions and the FE-evaluated commands. An admissible command is selected from this set by minimizing the loss. Acquired FE paths are retained for periodic updates of the neural operator and estimator.
}
\label{fig:workflow}
\end{figure}

This section defines the forward model that predicts the deformation of a magnetically active soft structure from known material properties, boundary conditions, and magnetic loading. The predicted response is used in the command-selection problem illustrated in \cref{fig:workflow}, where the objective is to select a magnetic field that achieves a desired response, characterized by quantities of interest (QoIs), while satisfying mechanical admissibility criteria. We first present the common mechanics formulation and then define the QoIs and command-selection problems for the two structures shown in \cref{fig:work}. The neural-operator approximation and error estimation are presented in \cref{sec:method}.

\subsection{Forward hMSM mechanics problem}

Let $\Omega\subset\mathbb R^2$ denote the reference domain of the structure, $\Gamma_D$ the boundary on which displacement is prescribed, and $\Gamma_N$ the loaded boundary. Both structures are modeled under plane strain, with in-plane displacement $\bu=(u_x,u_y)$ and unit out-of-plane stretch. The FE displacement space $\mathcal U$ incorporates the prescribed displacement conditions, while the corresponding test space $\mathcal U_0$ contains functions that vanish on $\Gamma_D$.

The known material and loading inputs are collected in $\bmpar\in\mathcal M$. Both examples use
\begin{equation}
\bmpar=(B_y,B_x,\gamma_G,f_x,f_y),
\label{eq:input}
\end{equation}
where $(B_x,B_y)$ are the components of the uniform applied magnetic flux density, $\gamma_G$ is a stiffness multiplier, and $(f_x,f_y)$ is the prescribed mechanical force. For each input $\bmpar$, the equilibrium problem seeks $\bu\in\mathcal U$ satisfying
\begin{equation}
\Rop(\bu;\bmpar)[\bm v]=\int_{\Omega}\bm P(\bu;\bmpar):\nabla\bm v\,\mathrm dA-\int_{\Gamma_N}\bar{\bm t}(\bmpar)\cdot\bm v\,\mathrm ds=0\qquad\forall\,\bm v\in\mathcal U_0,
\label{eq:weak_problem}
\end{equation}
where $\bm P$ is the in-plane first Piola--Kirchhoff stress and $\bar{\bm t}$ is the prescribed traction representing the force $(f_x,f_y)$ on $\Gamma_N$.

The stress in \cref{eq:weak_problem} is obtained from the elastic and magnetic energy of the material. The deformation gradient and its determinant are
\begin{equation}
\bm F=\bm I+\nabla\bu,\qquad J=\det\bm F,
\end{equation}
where $\bm I$ and $\bm F$ are $2\times2$ tensors, and the gradient is taken with respect to the reference coordinates. The total energy density and corresponding stress are
\begin{equation}
W(\bm F;\bmpar)=W_{\mathrm{el}}(\bm F;\gamma_G,\phi)+W_{\mathrm{mag}}(\bm F;\phi,\bm B^r,\bm B^a),\qquad \bm P=\frac{\partial W}{\partial\bm F}.
\end{equation}
The elastic and magnetic contributions are
\begin{align}
W_{\mathrm{el}}&=\frac{G}{2}\left[\operatorname{tr}(\bm F^T\bm F)-2-2\ln J\right]+\frac{K}{2}(J-1)^2,\label{eq:elastic_energy}\\
W_{\mathrm{mag}}&=-\frac{\phi}{\mu_0}(\bm F\bm B^r)\cdot\bm B^a,\qquad \bm B^a=(B_x,B_y).\label{eq:magnetic_energy}
\end{align}
Here, $\phi$ is the magnetic-particle volume fraction, $\bm B^r$ is the prescribed remanent flux density associated with the magnetic particles, $\bm B^a$ is the applied flux density, and $\mu_0=4\pi\times10^{-7}\,\mathrm{N/A^2}$ is the permeability of free space. The factor $\phi$ in $W_{\mathrm{mag}}$ accounts for the magnetic-particle volume fraction; it is not already included in $\bm B^r$. The effective shear modulus and volumetric coefficient are specified by
\begin{equation}
G=\gamma_G G_0\exp\left(\frac{2.5\phi}{1-1.35\phi}\right),\qquad K=500G,
\label{eq:effective_moduli}
\end{equation}
Here, $G_0$ is the reference matrix shear modulus. The elastic energy is the NH2 neo-Hookean form used in \cite{galloway_jha_hmsm2026}, and the effective shear modulus uses the Mooney relation that gave the best overall agreement among the relations tested there against mechanical stress--strain data. The magnetic term is the $\bm F$-based Zeeman energy used in \cite{zhao2019mechanics,galloway_jha_hmsm2026}. In the present calculations, $\gamma_G$ varies the stiffness with operating condition, and $K=500G$ specifies the volumetric coefficient. The geometry, material parameters, boundary conditions, and force application are specified for the platform and wishbone in \cref{sec:platform_setup,sec:wishbone_setup}, respectively.

For command selection, the input in \cref{eq:input} is separated into the operating condition $\bm\theta$ and magnetic command $\bm a$:
\begin{equation}
\bm\theta=(\gamma_G,f_x,f_y),\qquad \bm a=(B,\alpha).
\end{equation}
Here, $B$ is the magnitude of the applied magnetic field, and $\alpha$ is its angle measured from $+y$ toward $+x$. The field components are
\begin{equation}
B_x=B\sin\alpha,\qquad B_y=B\cos\alpha.
\label{eq:field_components}
\end{equation}
Thus, at a fixed operating condition, each candidate command specifies the applied field in \cref{eq:magnetic_energy} and defines one equilibrium problem through \cref{eq:input,eq:weak_problem}.

The nonlinear equilibrium problem is solved using 16 load increments. The applied magnetic field and mechanical force are increased proportionally from zero to their prescribed endpoint values, while $\gamma_G$ remains fixed. At each increment, Newton's method is initialized with the converged displacement from the preceding increment.
Both FE models are implemented in FEniCSx/DOLFINx \cite{baratta2023dolfinx}. In both models, the energy density enters the variational form directly, and the residual and tangent used by Newton's method are obtained by differentiation with respect to the displacement. No history-dependent constitutive update is performed at the material points.

Following the operator notation in \cite{jha2026mathematics}, this FE calculation defines the forward solution operator
\begin{equation}
\Fop:\mathcal M\rightarrow\mathcal U,\qquad \bu=\Fop(\bmpar).
\label{eq:forward_operator}
\end{equation}
For each prescribed input $\bmpar$, $\Fop(\bmpar)$ returns the converged displacement at the endpoint of the loading path. This displacement provides the QoIs used in the two command-selection problems defined below.

\subsection{Platform command problem}

\begin{figure}[h]
\centering
\includegraphics[width=0.68\textwidth]{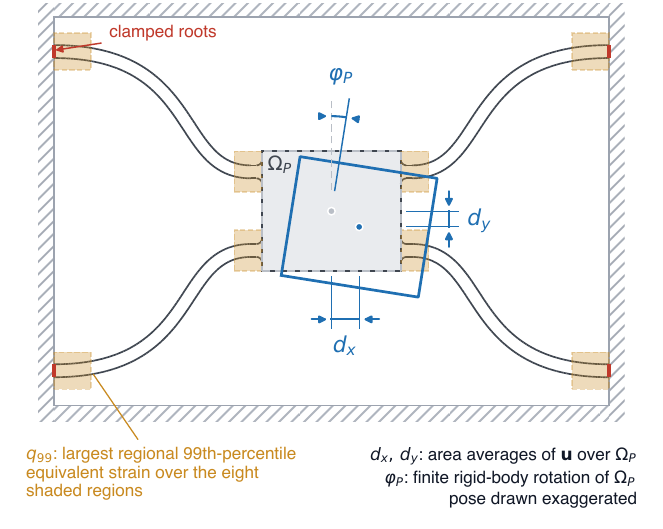}
\vspace{-10pt}
\caption{Platform quantities of interest. The translations $d_x$ and $d_y$ are area-averaged displacements of the stiff platform $\Omega_P$, and $\phi_P$ is its finite rigid-body rotation. The displaced outline is exaggerated to illustrate the pose quantities. The yellow regions mark the four ligament roots and four ligament--platform junctions; $q_{99}$ is the maximum of their area-weighted 99th-percentile equivalent strains.}
\label{fig:platform_qois}
\end{figure}

The primary example is the compliant positioning platform shown in \cref{fig:work}(a). For each operating condition $\bm\theta$ and candidate command $\bm a$, the displacement $\bu=\Fop(\bmpar)$ provides four QoIs describing the platform pose and local strain. Their geometric definitions are illustrated in \cref{fig:platform_qois}.

Let $\Omega_P$ denote the stiff central platform. Its translation is defined by the area-averaged displacement
\begin{equation}
\bm d=(d_x,d_y)=\frac{1}{|\Omega_P|}\int_{\Omega_P}\bu\,\mathrm dA.
\label{eq:platform_translation}
\end{equation}
The platform rotation $\phi_P$ is the angle of the rigid rotation that best aligns the reference and deformed platform coordinates after removing translation. This rotation is determined by minimizing the squared mismatch between corresponding positions.

Local strain is assessed in the eight regions near the four clamped roots and four ligament--platform junctions highlighted in \cref{fig:platform_qois}. Denote these regions by $\Omega_k^q$, $k=1,\ldots,8$. Let $\bm E$ be the three-dimensional extension of $(\bm F^T\bm F-\bm I)/2$ with $E_{33}=0$ under plane strain. The equivalent strain is $\varepsilon_{\mathrm{eq}}=\sqrt{(2/3)\operatorname{dev}\bm E:\operatorname{dev}\bm E}$, where the deviator is taken in three dimensions. The strain QoI is
\begin{equation}
q_{99}=\max_{1\leq k\leq8}\operatorname{perc}_{99}\left(\varepsilon_{\mathrm{eq}}\big|_{\Omega_k^q}\right).
\label{eq:platform_q99}
\end{equation}
Here, $\operatorname{perc}_{99}$ denotes the area-weighted 99th percentile: the strain level at or below which approximately $99\%$ of the region's area lies. The percentile is calculated separately within each region before the maximum over the eight regions is taken. This choice avoids basing admissibility on a single sampled peak while retaining the largest regional percentile among the roots and junctions.

The platform QoIs are collected as
\begin{equation}
\bq_P=\mathcal Q_P(\bu)=(d_x,\phi_P,d_y,q_{99}).
\label{eq:platform_qois}
\end{equation}
The horizontal translation $d_x$ and rotation $\phi_P$ are compared with the requested pose. The vertical drift $d_y$ and regional strain $q_{99}$ enter the mechanical admissibility checks.

The candidate command set is
\begin{equation}
\mathcal A_P={}\{(0,0)\}\cup\bigl\{(B,\alpha):B\in\{10,20,\ldots,80\}\ \si{mT}, \alpha\in\{-20,-10,0,10,20\}^{\circ}\bigr\},
\label{eq:platform_actions}
\end{equation}
which contains 41 commands, including a single zero-field command.

For a prescribed target pose $(d_x^*,\phi_P^*)$, the command loss is
\begin{equation}
L_P(\bm a;\bm\theta)=\left(\frac{d_x-d_x^*}{\SI{0.05}{mm}}\right)^2+\left(\frac{\operatorname{wrap}(\phi_P-\phi_P^*)}{0.2887^{\circ}}\right)^2.
\label{eq:platform_loss}
\end{equation}
The denominators set the relative scales of translation and rotation in the loss. The function $\operatorname{wrap}(\cdot)$ returns the signed angular difference on the principal interval.

Let $\mathcal A_P^{\mathrm{adm}}(\bm\theta)\subseteq\mathcal A_P$ contain the commands satisfying the prescribed mechanical admissibility criteria at operating condition $\bm\theta$. The reference command minimizes the loss over this admissible set:
\begin{equation}
\bm a_P^*(\bm\theta)=\underset{\bm a\in\mathcal A_P^{\mathrm{adm}}(\bm\theta)}{\operatorname{arg\,min}}\;L_P(\bm a;\bm\theta).
\label{eq:platform_command}
\end{equation}
The platform admissibility limits and operating conditions are given in \cref{sec:platform_setup,sec:platform_data}.

\subsection{Wishbone command problem}

\begin{figure}[h]
\centering
\includegraphics[width=0.7\textwidth]{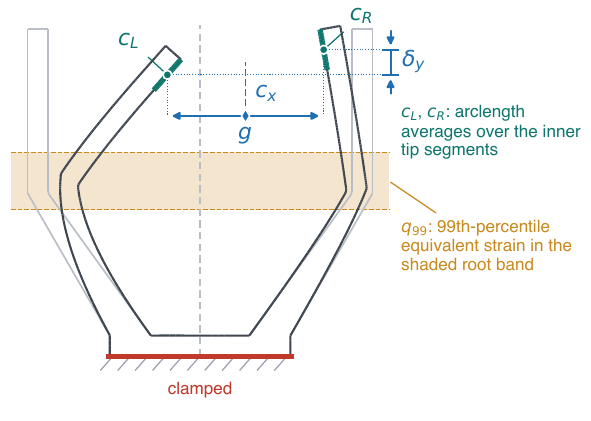}
\vspace{-10pt}
\caption{Wishbone quantities of interest. The points $\bm c_L$ and $\bm c_R$ represent averages of the deformed positions along the highlighted inner tip segments, weighted by reference arclength. Their horizontal separation defines the aperture $g$, the horizontal coordinate of their midpoint defines $c_x$, and their signed vertical difference defines $\delta_y$. The shaded root region $\Omega_W^q$ is used to calculate the area-weighted 99th-percentile equivalent strain $q_{99}$. The reference configuration is shown in light grey, and the deformed configuration illustrates the QoI definitions.}
\label{fig:wishbone_qois}
\end{figure}

The second example is the gripping wishbone shown in \cref{fig:work}(b), with two magnetically active fingers supported by passive lower branches. For each operating condition and candidate command, the displacement $\bu=\Fop(\bmpar)$ provides QoIs describing the relative positions of the finger tips and the strain near the root. These quantities are illustrated in \cref{fig:wishbone_qois}.

Let $\Gamma_L^c$ and $\Gamma_R^c$ denote the fixed inner tip segments in the reference configuration. Their averaged deformed positions are
\begin{equation}
\bm c_\nu=\frac{1}{|\Gamma_\nu^c|}\int_{\Gamma_\nu^c}(\bm X+\bu)\,\mathrm ds,\qquad \nu\in\{L,R\},
\label{eq:wishbone_tip_average}
\end{equation}
where $\mathrm ds$ and $|\Gamma_\nu^c|$ are measured in the reference configuration. From these averaged positions, define
\begin{equation}
g=c_{R,x}-c_{L,x},\qquad c_x=\frac{c_{R,x}+c_{L,x}}{2},\qquad \delta_y=c_{R,y}-c_{L,y}.
\label{eq:wishbone_displacement_qois}
\end{equation}
The aperture $g$ measures the horizontal separation of the two averaged tip positions, $c_x$ gives the horizontal coordinate of their midpoint, and $\delta_y$ measures their signed vertical mismatch.

The strain QoI uses the same area-weighted percentile as the platform but is evaluated over a single fixed root region $\Omega_W^q$, highlighted in \cref{fig:wishbone_qois}:
\begin{equation}
q_{99}=\operatorname{perc}_{99}\left(\varepsilon_{\mathrm{eq}}\big|_{\Omega_W^q}\right).
\label{eq:wishbone_q99}
\end{equation}
No maximum over multiple regions is required. The wishbone QoIs are collected as
\begin{equation}
\bq_W=\mathcal Q_W(\bu)=(g,c_x,\delta_y,q_{99}).
\label{eq:wishbone_qois}
\end{equation}

The candidate command set is
\begin{equation}
\mathcal A_W={}\{(0,0)\}\cup\bigl\{(B,\alpha):B\in\{20,40,60,80\}\ \si{mT}, \alpha\in\{-20,-10,0,10,20\}^{\circ}\bigr\},
\label{eq:wishbone_actions}
\end{equation}
which contains 21 commands, with all zero-field choices represented by $(0,0)$.

Eight fixed targets are generated at the reference operating condition
\begin{equation}
\bm\theta_{\mathrm{ref}}=(1,0,0)
\end{equation}
using magnetic commands not included in $\mathcal A_W$. Each target $j$ specifies a desired aperture and lateral midpoint, $(g_j^*,c_{x,j}^*)$.

For target $j$, the command loss is
\begin{equation}
\begin{split}
L_{W,j}(\bm a;\bm\theta)={}&\left(\frac{g-g_j^*}{\SI{5}{mm}}\right)^2+\left(\frac{c_x-c_{x,j}^*}{\SI{3}{mm}}\right)^2\\
&+0.25\left(\frac{\delta_y}{\SI{3}{mm}}\right)^2+0.002\left(\frac{B}{\SI{80}{mT}}\right)^2.
\end{split}
\label{eq:wishbone_loss}
\end{equation}
The first two terms measure the difference from the requested tip configuration, the third penalizes vertical mismatch, and the final term weakly penalizes magnetic-field magnitude. The aperture and root strain also enter the mechanical admissibility checks, whose limits are given in \cref{sec:wishbone_setup}.

Let $\mathcal A_W^{\mathrm{adm}}(\bm\theta)\subseteq\mathcal A_W$ contain the mechanically admissible commands at operating condition $\bm\theta$. The reference command for target $j$ is
\begin{equation}
\bm a_{W,j}^*(\bm\theta)=\underset{\bm a\in\mathcal A_W^{\mathrm{adm}}(\bm\theta)}{\operatorname{arg\,min}}\;L_{W,j}(\bm a;\bm\theta).
\label{eq:wishbone_command}
\end{equation}

Both examples use the displacement field to evaluate the achieved response and mechanical admissibility before selecting the command that minimizes the loss over the admissible set. \cref{sec:method} introduces the neural-operator approximation and selective FE evaluations used to perform this calculation, together with the reuse of FE loading paths for adaptation.

\section{Neural operator, QoI-error estimator, and adaptation}
\label{sec:method}

The command problems in \cref{sec:problem} require displacement predictions for a discrete set of candidate magnetic commands. The procedure in \cref{fig:workflow} combines neural-operator predictions with a fixed number of FE evaluations. For each operating condition, the neural operator predicts the displacement for every candidate command, and an empirical error estimator determines which predicted QoIs may be used. Four commands are also evaluated by FE. Command selection then uses the accepted surrogate predictions and the FE results, while the complete FE loading paths are retained for subsequent adaptation.

\subsection{POD neural operator}

We use a projection-based neural operator \cite{BhattacharyaHosseiniKovachkiEtAl2020} (see \cite[Sections 2.3 and 4.2]{jha2026mathematics}) to approximate the forward map $\bmpar\mapsto\bu$ in \cref{eq:forward_operator} by learning the coefficients of a reduced displacement representation. In this section, $\bu\in\mathbb R^{p_U}$ denotes the vector of FE displacement degrees of freedom. A proper orthogonal decomposition (POD) gives
\begin{equation}
\bu\approx\overline{\bu}+\bm V_r\widetilde{\bu},\qquad \bm V_r^T\bm M\bm V_r=\bm I,\qquad r=16,
\label{eq:pod}
\end{equation}
where $\overline{\bu}$ is the mean of the FE displacement snapshots, $\bm M$ is the displacement mass matrix, $\bm V_r\in\mathbb R^{p_U\times r}$ contains the retained POD modes, and $\widetilde{\bu}\in\mathbb R^r$ contains their coefficients. Each snapshot is an FE displacement solution at a loading increment. The basis is orthonormal with respect to the mass inner product, with associated norm $\|\bm v\|_{\bm M}=(\bm v^T\bm M\bm v)^{1/2}$.

The learned forward map is
\begin{equation}
\FNOp(\bmpar;\bm\Theta)=\overline{\bu}+\bm V_r\widetilde{\Fop}_{\mathrm{NOp}}\bigl(\bm\xi(\bmpar);\bm\Theta\bigr),
\label{eq:nop}
\end{equation}
where $\widetilde{\Fop}_{\mathrm{NOp}}$ is the neural network that predicts the POD coefficients, $\bm\Theta$ contains its trainable parameters, and $\bm\xi(\bmpar)$ is its input vector. This vector contains the five components of $\bmpar$ in \cref{eq:input}, augmented by
\begin{equation}
\frac{B_y}{\gamma_G},\qquad \frac{B_x}{\gamma_G},\qquad \frac{f_x}{\gamma_G},\qquad \frac{f_y}{\gamma_G},
\end{equation}
giving nine inputs. All nine quantities are then affinely scaled using fixed bounds determined from the nominal input box; inputs outside those bounds are not clipped. The additional ratios express the applied field and mechanical force relative to the stiffness multiplier.

The data and training settings for the initial operator $M_0$ are summarized in \cref{tab:initial_training}. For each example, disjoint nominal FE loading paths are used for training, validation, and final evaluation. Validation determines which trained operator is retained, while the test paths are reserved for final assessment. The platform and wishbone use the same settings but have separate POD bases, neural operators, estimators, and data.

\begin{table}[h]
\centering
\caption{Data, coefficient-network architecture, and initial training settings for each example.}
\label{tab:initial_training}
\small
\renewcommand{\arraystretch}{1.08}
\begin{tabular}{ll}
\toprule
\textbf{Quantity} & \textbf{Specification} \\
\midrule
\multicolumn{2}{@{}l}{\textit{Nominal FE data}} \\
Training paths & 200 \\
Validation paths & 25 \\
Held-out test paths & 30 \\
Retained increments per training path & 16 \\
\midrule
\multicolumn{2}{@{}l}{\textit{Coefficient network}} \\
Architecture & Fully connected neural network \\
Hidden layers & Four layers, each containing 256 units \\
Activation & Gaussian error linear unit (GELU) \\
Outputs & 16 POD coefficients \\
Trainable parameters & 204,048 \\
\midrule
\multicolumn{2}{@{}l}{\textit{Initial training}} \\
Optimizer & AdamW \\
Learning rate & $10^{-3}$ \\
Weight decay & $10^{-6}$ \\
Batch size & 128 \\
Maximum training duration & 4000 epochs \\
Early stopping & 250 consecutive epochs without improvement \\
\bottomrule
\end{tabular}
\end{table}

Training uses AdamW, an adaptive gradient optimizer with weight decay. The squared coefficient error corresponds to the squared mass-norm displacement error within the retained POD space; it does not include the error from truncating the POD representation.

\subsection{QoI error and empirical estimate}

The accuracy needed for command selection is assessed through the QoIs defined in \cref{sec:problem}. For an input $\bmpar$, the current neural operator gives $\widehat{\bu}=\FNOp(\bmpar;\bm\Theta)$ and predicted QoIs $\widehat{\bq}=\mathcal Q(\widehat{\bu})$. When the FE solution is available, these predictions can be compared with $\bq=\mathcal Q(\bu)$ using
\begin{equation}
\EQ=\max_{1\leq k\leq4}\frac{|\widehat q_k-q_k|}{\tau_k},
\label{eq:true_error}
\end{equation}
where $\tau_k$ is the prescribed error tolerance for the $k$th QoI. Thus, $\EQ\leq1$ means that all four QoI errors satisfy their error tolerances.

For the platform, the error tolerances are
\begin{equation}
\bm\tau_P=\left(\SI{0.05}{mm},0.2887^{\circ},\SI{0.05}{mm},0.01\right),
\label{eq:platform_qoi_tolerances}
\end{equation}
corresponding to $(d_x,\phi_P,d_y,q_{99})$. For the wishbone,
\begin{equation}
\bm\tau_W=\left(\SI{0.05}{mm},\SI{0.05}{mm},\SI{0.05}{mm},0.01\right),
\label{eq:wishbone_qoi_tolerances}
\end{equation}
corresponding to $(g,c_x,\delta_y,q_{99})$. These tolerances constrain the surrogate error, whereas the mechanical admissibility limits constrain the response itself.

At an input not evaluated by FE, $\EQ$ is unavailable. Its empirical estimate is constructed from four scalar features: two describe the residual and linearized response at the predicted displacement, and two describe the location of the input relative to the available data.

For the first two features, the nonlinear residual is assembled at $\widehat{\bu}$ and the linear system
\begin{equation}
\bm A(\widehat{\bu};\bmpar)\,\delta\bu=-\bm R(\widehat{\bu};\bmpar),\qquad \bm A(\widehat{\bu};\bmpar)=D_{\bu}\bm R(\widehat{\bu};\bmpar)
\label{eq:linearized_indicator}
\end{equation}
is solved once. Here, $\bm R$ is the assembled residual of the equilibrium problem in \cref{eq:weak_problem}, and $\bm A$ is its derivative with respect to the displacement vector. The resulting $\delta\bu$ describes the response of this linearized problem to the residual. It is used only for error estimation; the QoIs continue to be evaluated from $\widehat{\bu}$, not from $\widehat{\bu}+\delta\bu$.

The two residual-based features are
\begin{equation}
\rho=\frac{\|\bm R(\widehat{\bu};\bmpar)\|_2}{\|\bm R(\bm 0;\bmpar)\|_2},\qquad \eta_{\delta u}=\frac{\|\delta\bu\|_{\bm M}}{\|\widehat{\bu}\|_{\bm M}}.
\label{eq:residual_features}
\end{equation}
The quantity $\rho$ measures the residual at the predicted displacement relative to that at zero displacement, while $\eta_{\delta u}$ measures the size of the linearized response relative to the prediction.

The distance features use the five-component input vector in \cref{eq:input}. Let $\bm\ell$ and $\bm h$ denote the componentwise lower and upper bounds of the nominal input box, and let $\bm w=\bm h-\bm\ell$. The normalized distance outside this box is
\begin{equation}
d_{\mathrm{box}}=\left\|\frac{[\bm\ell-\bmpar]_+ + [\bmpar-\bm h]_+}{\bm w}\right\|_2,
\label{eq:box_distance}
\end{equation}
where $[\cdot]_+$ replaces negative components by zero and division by $\bm w$ is componentwise. Thus, $d_{\mathrm{box}}=0$ inside the nominal box. The distance to the nearest input in the augmented data is
\begin{equation}
d_{\mathrm{aug}}=\min_{\bmpar_i\in\mathcal D_{\mathrm{aug}}}\left\|\frac{\bmpar-\bmpar_i}{\bm w}\right\|_2,
\label{eq:aug_distance}
\end{equation}
where $\mathcal D_{\mathrm{aug}}$ contains the increment inputs from the 200 nominal training paths and the FE paths acquired during adaptation. For a fixed input, $d_{\mathrm{box}}$ remains unchanged as data are added, whereas $d_{\mathrm{aug}}$ reflects the additional inputs represented by the acquired paths.

The estimator uses the feature vector
\begin{equation}
\bm s=\left(\log\rho,\log\eta_{\delta u},d_{\mathrm{box}},d_{\mathrm{aug}}\right).
\label{eq:features}
\end{equation}
Before taking logarithms, values of $\rho$ and $\eta_{\delta u}$ below $10^{-16}$ are replaced by $10^{-16}$. The same lower bound is applied to $\EQ$ when fitting $\log\EQ$.

The estimator is fitted using development data with FE reference solutions, divided into regression and calibration subsets. The regression subset supplies the componentwise feature means $\bm\mu_s$ and standard deviations $\bm\sigma_s$ used for standardization:
\begin{equation}
\bm z=\operatorname{diag}(\bm\sigma_s)^{-1}(\bm s-\bm\mu_s).
\label{eq:standardized_features}
\end{equation}
Ridge regression fits a linear approximation $\beta_0+\bm\beta^T\bm z$ to $\log\EQ$, with a quadratic penalty on the regression coefficients. Here, $\beta_0$ is the intercept and $\bm\beta$ contains the coefficients. The penalty parameter is selected from $10^{-4},10^{-3},\ldots,10^2$ using leave-path-out validation, in which all samples belonging to a validation path are excluded from the regression fit.

The calibration subset supplies an additive adjustment to the fitted log-error:
\begin{equation}
\begin{aligned}
\Ehat&=\exp\left(\beta_0+\bm\beta^T\bm z+q_{\mathrm{cal}}\right),\\
q_{\mathrm{cal}}&=Q_{0.90}\left[\log\EQ-(\beta_0+\bm\beta^T\bm z)\right].
\end{aligned}
\label{eq:estimator}
\end{equation}
Here, $Q_{0.90}$ is the empirical 90th percentile of the regression residuals on the calibration subset. Exponentiation returns the estimate to the scale of $\EQ$.

A surrogate prediction is accepted for command selection when $\Ehat\leq1$ and the predicted field satisfies the prescribed mechanical admissibility checks. Values of $\log\Ehat$ above $\log(10^6)$ are capped and recorded, and the corresponding predictions are not accepted. A false acceptance occurs when an accepted prediction is subsequently evaluated by FE and found to have $\EQ>1$. This acceptance criterion is empirical: calibration does not establish a guaranteed upper bound on the error at an individual input.

\subsection{Selective FE evaluation and command calculation}

For a fixed operating condition $\bm\theta$, the current neural operator and estimator evaluate every command in the candidate set. Exactly four commands are selected for FE evaluation under either an estimator-guided or a random acquisition strategy. The number of FE evaluations is fixed, regardless of how many surrogate predictions are accepted.

Estimator-guided acquisition first considers commands whose surrogate predictions are not accepted. If this set is nonempty, the command with the largest $\Ehat$ is selected first. Further commands from this set are chosen to maximize their minimum normalized input-space distance from those already selected, with $\Ehat$ used to break ties. This rule prioritizes commands with unaccepted predictions while separating the selected inputs. If fewer than four predictions are unaccepted, the remaining evaluations are assigned to the unselected commands with the largest $\Ehat$.

Random acquisition selects four commands uniformly without replacement from the complete candidate set using fixed random seeds. This comparison uses the same FE budget without using the error estimate to determine which commands are evaluated.

After these evaluations, the available command set is
\begin{equation}
\mathcal A_{\mathrm{avail}}=\mathcal A_{\mathrm{FE}}\cup\mathcal A_{\mathrm{NOp}}^{\mathrm{acc}},
\label{eq:available_actions}
\end{equation}
where $\mathcal A_{\mathrm{FE}}$ contains the four FE-evaluated commands and $\mathcal A_{\mathrm{NOp}}^{\mathrm{acc}}$ contains the commands with accepted surrogate predictions that were not selected for FE. For each command in $\mathcal A_{\mathrm{FE}}$, the FE QoIs replace the surrogate values. A command whose prediction is not accepted and that is not selected for FE remains unavailable.

The available QoIs enter the loss functions in \cref{eq:platform_loss,eq:wishbone_loss}. The \emph{four-FE calculation} selects among the four FE-evaluated commands, whereas the \emph{hybrid calculation} also includes commands with accepted neural-operator predictions. In both calculations, mechanical admissibility is assessed using the corresponding FE solution or surrogate prediction, and the command with the smallest loss among the admissible candidates is selected.

For final assessment, the complete candidate command set at each test condition is also evaluated by FE. These reference evaluations provide the FE-optimal command and are kept separate from command selection, estimator fitting, and adaptation. Command quality is measured by agreement with the FE-optimal command and by the FE-evaluated regret,
\begin{equation}
R=L_{\mathrm{FE}}\left(\bm a_{\mathrm{selected}}\right)-L_{\mathrm{FE}}\left(\bm a_{\mathrm{FE}}^*\right),
\label{eq:regret}
\end{equation}
where $\bm a_{\mathrm{FE}}^*$ minimizes the loss over the complete FE-admissible command set. Both losses are evaluated for the same operating condition and target. Regret therefore measures the excess FE loss of the selected command relative to the reference optimum; $R=0$ when the two losses are equal.

\subsection{Periodic adaptation}

The FE solutions at the loading-path endpoints provide the QoIs used for command selection. The intermediate equilibrium solutions are also retained, so each acquired path supplies 16 displacement samples for adaptation. Four paths are collected at each of four adaptation conditions. Updates after eight and 16 acquired paths produce candidate operators $M_1$ and $M_2$, respectively. The POD mean and basis remain fixed; only the learned input-to-coefficient map is updated. A candidate update is used only if it satisfies the validation requirements described below.

At each adaptation condition, the acquired path with the smallest command index is reserved for adaptation validation. The remaining acquired paths are used for continued training. Each training batch contains equal numbers of acquired samples and samples from 32 fixed nominal paths. Reusing the nominal samples alongside the new data is intended to preserve accuracy within the initial training regime. All 16 increments from each path are retained.

Continued training starts from the parameters of the current operator and uses AdamW with a learning rate of $10^{-4}$ and a batch size of 128. Training continues for at most 1000 epochs and stops after 100 epochs without improvement. A candidate updated operator must reduce the adaptation-validation mass error relative to the operator before updating. It must also satisfy two requirements on the 25 nominal validation paths:
\begin{equation}
\operatorname{median}_{\mathrm{val}}\EQ\leq0.25,\qquad \#\{\EQ\leq1\}\geq24.
\label{eq:update_acceptance}
\end{equation}
Among the candidates satisfying these requirements, the operator with the smallest adaptation-validation mass error is retained. If none qualifies, the preceding operator remains in use. Preservation of nominal accuracy is subsequently assessed on the separate 30-path test set, not on the validation data used to select the update.

The empirical error estimator is refitted for each accepted updated operator using its predictions and the available FE data. Separate estimator fits are used for acquisition and final assessment. The acquisition estimator uses only FE reference data available before the four new commands are selected at the current adaptation condition. The estimator used for final assessment may additionally use the prescribed development data and previously acquired paths, but neither fit uses the held-out shifted paths or final command-test data. These final data therefore assess the resulting predictions and decisions without influencing the estimator used to make them.

\section{Compliant positioning platform}
\label{sec:platform}

The platform study examines whether FE loading paths collected under changing stiffness and mechanical loading can improve subsequent displacement predictions and positioning decisions. The command problem uses horizontal translation and rotation to specify the requested pose, with vertical drift and regional strain entering the admissibility checks; see \cref{eq:platform_qois,eq:platform_loss}. The numerical setup is presented first, followed by the results for forward accuracy, error estimation, command selection, and computational work.

\subsection{Numerical setup}
\label{sec:platform_setup}

\subsubsection{Geometry and FE model}

\begin{figure}[h!]
\centering
\includegraphics[width=\linewidth]{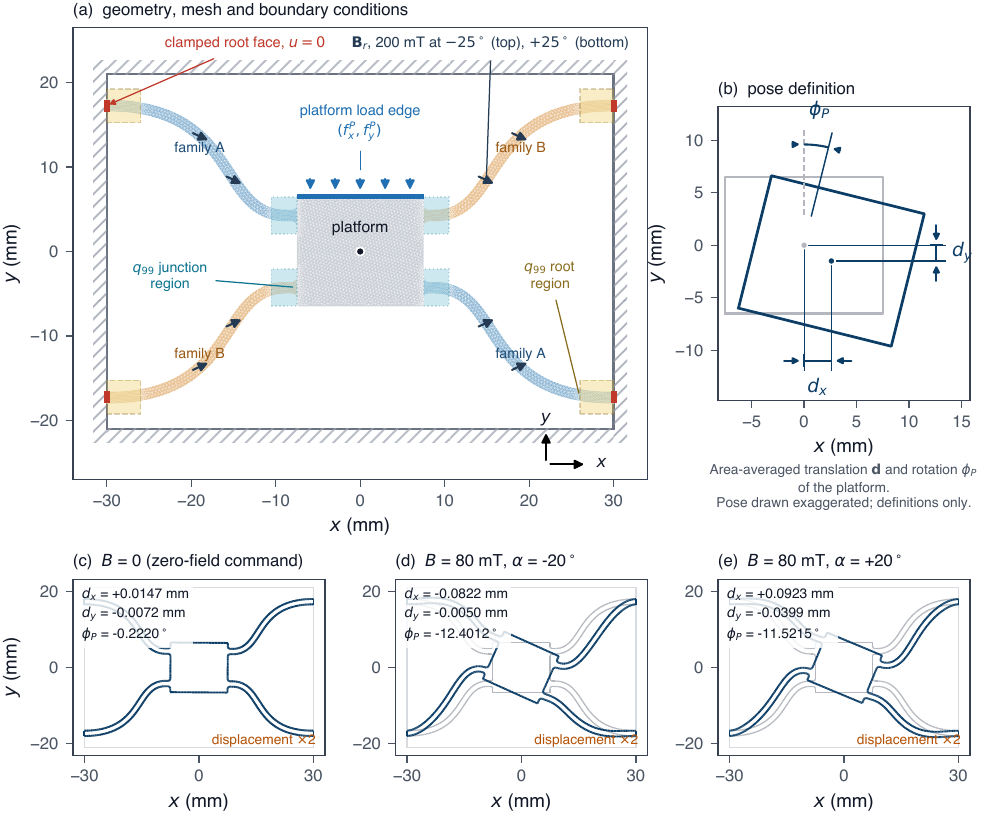}
\caption{Platform geometry and representative FE responses. (a) Reference geometry, mesh, clamped roots, and loaded platform edge. The yellow and cyan regions identify the root and junction regions used to calculate $q_{99}$. (b) Platform translation and rotation. (c)--(e) FE deformations at $T_1$ in \cref{tab:platform_conditions}, for $B=0$ and for $B=\SI{80}{mT}$ at $\alpha=-20^{\circ}$ and $+20^{\circ}$, respectively. Displacements in the deformation panels are amplified by a factor of two.}
\label{fig:platform_setup}
\end{figure}

The platform geometry, boundary conditions, mesh, and representative deformations are shown in \cref{fig:platform_setup}. Four curved hMSM ligaments of width \SI{1.4}{mm} support a stiff central platform. The outer ligament roots are clamped, and the prescribed force $(f_x^P,f_y^P)$ is applied to the top edge of the platform. The modeled thickness is \SI{3}{mm}. The base ligament shear modulus is \SI{100}{kPa}, and the platform stiffness is 100 times the ligament stiffness. The ligaments have a magnetic-particle volume fraction $\phi=0.30$ and a remanent flux density of magnitude \SI{200}{mT}, directed at $-25^{\circ}$ in the upper ligaments and $+25^{\circ}$ in the lower ligaments. The platform is magnetically inactive.

The domain is discretized using 8,360 triangles and vector-valued quadratic displacement elements, giving 35,358 displacement degrees of freedom. The equilibrium solution follows the Newton iteration and 16-increment loading procedure in \cref{sec:problem}. An FE path is admissible only if the nonlinear solution converges, $J$ is positive at the sampled locations, and no contact or self-intersection is detected. The platform must remain separated from the frame and the central cavity must remain open. The response must also satisfy $|d_y|\leq\SI{0.05}{mm}$, $q_{99}\leq0.10$, and $\varepsilon_{\mathrm{rig}}<10^{-3}$. Here, $\varepsilon_{\mathrm{rig}}=\|\bu-\bu_{\mathrm{rig}}\|_{L^2(\Omega_P)}/\|\bu\|_{L^2(\Omega_P)}$, where $\bu_{\mathrm{rig}}(\bm X)=\bm d+[\bm Q_P(\phi_P)-\bm I](\bm X-\bm X_c)$. The matrix $\bm Q_P(\phi_P)$ is the planar rotation through $\phi_P$, and $\bm X_c$ is the reference centroid of the platform.

\subsubsection{Nominal data and shifted operating conditions}
\label{sec:platform_data}

Nominal FE loading paths are generated from endpoint inputs in the ranges
\begin{equation}
\begin{aligned}
B&\in[0,80]\ \si{mT},\qquad \alpha\in[-20,20]^{\circ},\qquad \gamma_G\in[0.95,1.05],\\
f_x^P&\in[-0.10,0.10]\ \si{mN},\qquad f_y^P\in[-0.65,0]\ \si{mN}.
\end{aligned}
\label{eq:platform_nominal_box}
\end{equation}
The 200 training, 25 validation, and 30 test paths are disjoint, as specified in \cref{tab:initial_training}. All 16 increments of each training path are retained. The 41 candidate commands in \cref{eq:platform_actions} lie within the magnetic-field ranges used to generate the nominal data.

The controlled shift reduces the stiffness multiplier and increases the horizontal force while preserving the vertical-force and magnetic-field ranges. Three shifted input ranges, S1, S2, and S3, are prescribed before final evaluation and assessed using 12 preliminary paths each; see \cref{tab:platform_shift}. The range selected for the final shifted test is the first for which the initial operator has median $\EQ>1$ on these preliminary paths. S3 meets this criterion with median $\EQ=1.181$. A separate set of 30 S3 paths is reserved before adaptation for final evaluation.

\begin{table}[h]
\centering
\caption{Nominal and shifted input ranges for the platform. The magnetic-field ranges are those in \cref{eq:platform_nominal_box}, and $f_y^P\in[-0.65,0]\ \si{mN}$ for every set. Each shifted range is assessed using 12 preliminary paths.}
\label{tab:platform_shift}
\begin{tabular}{@{}lccc@{}}
\toprule
Input range & $\gamma_G$ & $f_x^P$ (mN) & Preliminary median $\EQ$ \\
\midrule
Nominal & $[0.95,1.05]$ & $[-0.10,0.10]$ & \textemdash\\
S1 & $[0.86,0.94]$ & $[0.25,0.70]$ & 0.275\\
S2 & $[0.78,0.89]$ & $[0.45,1.00]$ & 0.735\\
S3 & $[0.72,0.84]$ & $[0.70,1.30]$ & 1.181\\
\bottomrule
\end{tabular}
\end{table}

Four operating conditions, $A_1$ through $A_4$, are used to acquire FE paths for adaptation. Four additional conditions, $T_1$ through $T_4$, are reserved for testing command selection. Their inputs are listed in \cref{tab:platform_conditions}. Eight target poses are generated from FE solutions at the nominal reference condition $(\gamma_G,f_x^P,f_y^P)=(1,0,\SI{-0.5}{mN})$: two use $B=\SI{30}{mT}$ with $\alpha=\pm15^{\circ}$, and six use $B\in\{50,70\}\ \si{mT}$ with $\alpha\in\{-15,0,15\}^{\circ}$. Six target-generating commands therefore use angles between those in the candidate set, while the two zero-angle commands belong to the set. The target poses remain fixed across the four test conditions.

\begin{table}[h]
\centering
\caption{Platform operating conditions. Force components are in mN. Four FE paths are acquired at each $A_k$. The conditions $T_1$ through $T_4$ are reserved for testing command selection.}
\label{tab:platform_conditions}
\small
\begin{tabular}{@{}lrrr@{\qquad}lrrr@{}}
\toprule
\multicolumn{4}{c}{Adaptation} & \multicolumn{4}{c}{Command tests}\\
\cmidrule(lr){1-4}\cmidrule(lr){5-8}
Condition & $\gamma_G$ & $f_x^P$ & $f_y^P$ & Condition & $\gamma_G$ & $f_x^P$ & $f_y^P$\\
\midrule
$A_1$ & 0.816 & 0.82 & -0.4875 & $T_1$ & 0.804 & 0.88 & -0.260\\
$A_2$ & 0.792 & 0.94 & -0.1625 & $T_2$ & 0.780 & 1.00 & -0.520\\
$A_3$ & 0.768 & 1.06 & -0.3250 & $T_3$ & 0.756 & 1.12 & -0.130\\
$A_4$ & 0.744 & 1.18 & -0.4875 & $T_4$ & 0.732 & 1.24 & -0.390\\
\bottomrule
\end{tabular}
\end{table}

\subsubsection{Training, adaptation, and evaluation}
\label{sec:platform_evaluation}

The initial operator $M_0$ uses the rank-16 POD basis, coefficient network, and training procedure in \cref{sec:method}. The initial error estimator uses 32 FE endpoint solutions: 16 nominal and 16 shifted. Eight nominal and eight shifted endpoints form the regression subset, and the remaining 16 form the calibration subset used for the quantile adjustment in \cref{eq:estimator}. Subsequent estimator fits use these original endpoints together with those acquired in the corresponding adaptation sequence. Acquired endpoints from $A_1$ and $A_3$ are assigned to regression, while those from $A_2$ and $A_4$ are assigned to calibration.

Five paired adaptation sequences compare random and estimator-guided acquisition. Each sequence collects four FE paths at each $A_k$, giving eight paths after $A_2$ and 16 after $A_4$. These data are used to construct candidate operators $M_1$ and $M_2$. An update is used for subsequent command calculations only if it satisfies \cref{eq:update_acceptance}; otherwise, the preceding operator remains in use.

Forward accuracy is assessed using $\EQ$ in \cref{eq:true_error} on the 30 held-out nominal paths and the 30 held-out S3 paths. The condition $\EQ\leq1$ means that all four QoI errors satisfy their prescribed tolerances. Estimator reliability is assessed by comparing $\Ehat$ with $\EQ$, including false acceptance of inaccurate predictions, false rejection of accurate predictions, and their rank correlation. For the field comparisons, each 30-path test set is ordered by increasing $\EQ$ for the initial operator, and the 15th path is shown. This selects the lower of the two middle paths rather than the path with the smallest error.

Command selection is assessed for the eight fixed targets at each of the four test conditions in \cref{tab:platform_conditions}, giving 32 combinations of target and operating condition per operator. The complete 41-command FE response set at each condition provides the reference optimum and the losses used to calculate regret in \cref{eq:regret}. Each hybrid calculation uses only its four selected FE evaluations and the accepted surrogate predictions, as described in \cref{sec:method}. The remaining reference FE solutions are used for assessment, not to select the command or update the operator and estimator.

\FloatBarrier
\subsection{Results}
\label{sec:platform_results}

\subsubsection{Nominal forward accuracy}

Within the nominal input range in \cref{eq:platform_nominal_box}, the initial operator $M_0$ accurately reproduces the FE response. Its median $\EQ$ is 0.0088 on the 25 validation paths and 0.0081 on the 30 held-out test paths. Every prediction in both sets satisfies $\EQ\leq1$.

\begin{figure}[h]
\centering
\includegraphics[width=\linewidth]{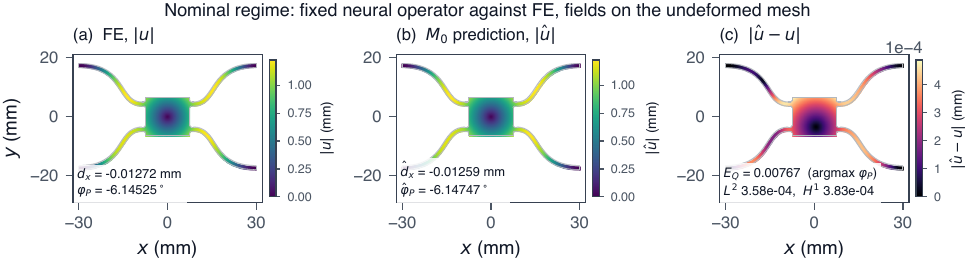}
\caption{Platform FE solution and initial neural-operator prediction for the 15th of 30 nominal test paths ordered by the initial operator's $\EQ$. The first two panels use the same displacement-magnitude scale; the third shows the pointwise displacement error. All fields are shown on the reference configuration.}
\label{fig:platform_nominal_field}
\end{figure}

The field comparison in \cref{fig:platform_nominal_field} uses the 15th path after ordering the 30 nominal test paths by the initial operator's error, as specified in \cref{sec:platform_evaluation}. For this path, the relative mass-weighted displacement error is $3.58\times10^{-4}$. The FE pose is $(d_x,\phi_P)=(-0.0127\ \mathrm{mm},-6.1452^{\circ})$, compared with $(-0.0126\ \mathrm{mm},-6.1475^{\circ})$ from $M_0$. The corresponding QoI error is $\EQ=0.0077$.

\subsubsection{Shifted forward accuracy and adaptation}

The same operator loses accuracy in S3, the reduced-stiffness and increased-horizontal-force range selected in \cref{tab:platform_shift}. On the separate 30-path S3 test, median $\EQ$ increases to 1.147, and 21 predictions exceed the prescribed tolerance. The median relative mass-weighted displacement error is 0.0431.

To distinguish error in the learned coefficients from error due to the reduced representation, the same 30 FE displacement fields are projected onto the fixed rank-16 POD basis. The QoIs computed from these projected fields have median $\EQ=7.08\times10^{-4}$. The basis therefore represents the tested shifted responses accurately; the substantial error of $M_0$ is primarily associated with the learned input-to-coefficient map.

\begin{figure}[h]
\centering
\includegraphics[width=\linewidth]{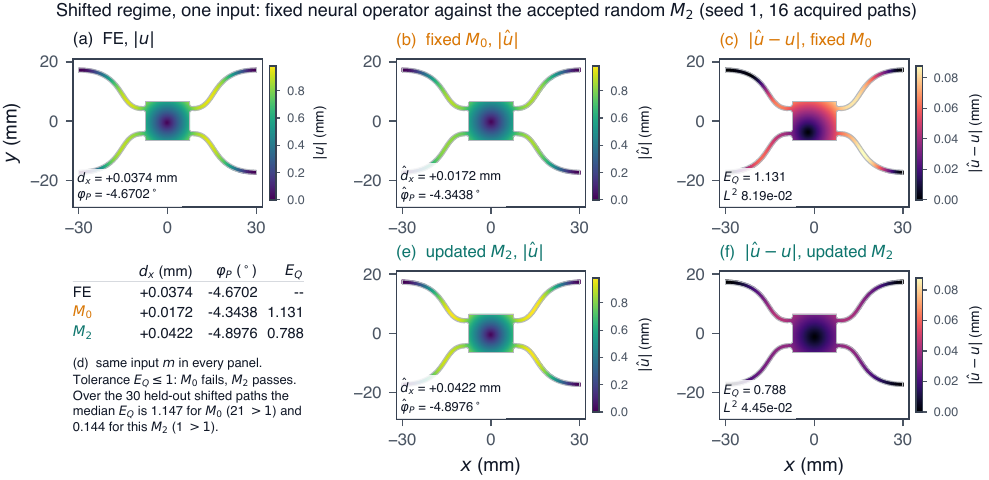}
\caption{Platform FE solution, initial prediction, and accepted 16-path updated prediction for the same S3 test input. The input is the 15th of 30 paths ordered by the initial operator's $\EQ$. The update uses random-acquisition seed 1. Displacement panels share one color scale, and error panels share a separate scale. Results across all five acquisition sequences are given in \cref{fig:platform_forward}.}
\label{fig:platform_ood_field}
\end{figure}

\cref{fig:platform_ood_field} compares the initial and updated predictions for the 15th path after ordering the S3 test set by the initial operator's error. The FE response is $d_x=\SI{0.0374}{mm}$ and $\phi_P=-4.6702^{\circ}$. The initial operator predicts $d_x=\SI{0.0172}{mm}$ and $\phi_P=-4.3438^{\circ}$, giving $\EQ=1.131$. The accepted $M_2$ operator from random-acquisition seed 1 predicts $d_x=\SI{0.0422}{mm}$ and $\phi_P=-4.8976^{\circ}$, reducing the error to $\EQ=0.788$. This comparison uses the same input before and after adaptation; it is not selected by the updated operator's error.

\begin{figure}[h]
\centering
\includegraphics[width=\linewidth]{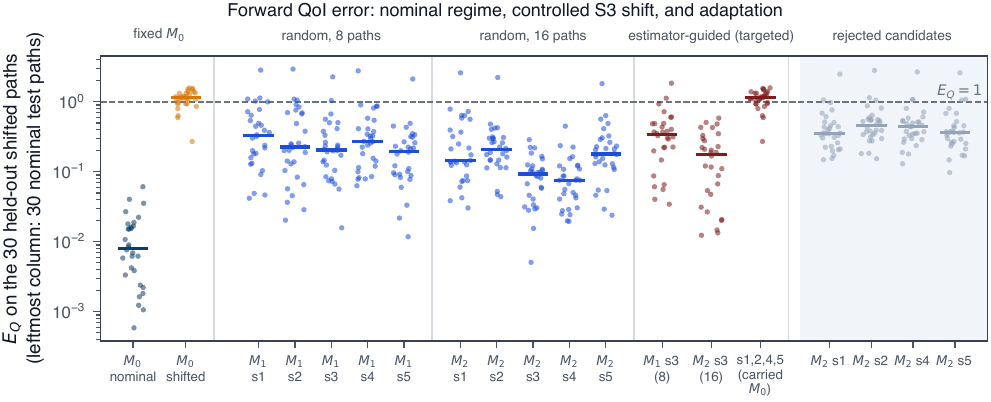}
\caption{Platform QoI errors before and after adaptation. The leftmost distribution uses the 30 nominal test paths; the remaining distributions use the 30 S3 test paths from \cref{tab:platform_shift}. The five random-acquisition sequences are shown individually. Only estimator-guided seed 3 satisfies the update-acceptance requirements at both stages. The shaded candidate updates reduce shifted error but are not used for command selection because they fail validation.}
\label{fig:platform_forward}
\end{figure}

\cref{fig:platform_forward} reports the errors over the complete S3 test set. For each random-acquisition sequence, the median is calculated over the 30 test paths. The median of these five sequence-level medians decreases from 1.147 for $M_0$ to 0.226 after eight acquired paths and 0.144 after 16 paths. All five random-acquisition sequences produce accepted updates at both stages under \cref{eq:update_acceptance}. After 16 paths, the numbers of predictions with $\EQ>1$ are 1, 1, 0, 0, and 1 across the five sequences.

For estimator-guided acquisition, only seed 3 produces accepted updates at both stages. Its 16-path operator has median $\EQ=0.178$, with all 30 shifted predictions satisfying the tolerance. The other four candidate $M_2$ operators reduce the shifted median to values between 0.35 and 0.46 but fail the adaptation-validation requirement. They are shown for comparison and are not used for command selection. Every accepted update preserves $\EQ\leq1$ on all 30 nominal test paths.

The different update-acceptance counts do not by themselves establish that random acquisition is preferable. Estimator-guided acquisition selects the zero-field command first at each adaptation condition. Under the smallest-command-index rule in \cref{sec:method}, zero-field paths therefore supply the adaptation-validation data. Four estimator-guided sequences share one 16-path set, while seed 3 selects different commands at $A_3$ and $A_4$ and produces a second set. Random acquisition produces five distinct path sets. The comparison consequently depends on both the acquired data and the validation requirement and does not support a general ranking of the acquisition strategies.

\subsubsection{QoI-error estimation}

\begin{figure}[h]
\centering
\includegraphics[width=0.55\linewidth]{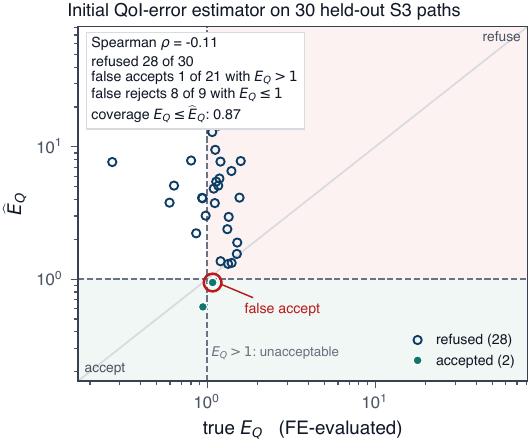}
\caption{Initial platform error estimate versus FE-evaluated QoI error on the 30 S3 test paths. Acceptance requires $\Ehat\leq1$, while the actual accuracy requirement is $\EQ\leq1$. The estimator accepts one inaccurate prediction and rejects eight accurate predictions. Its rejection of 28 of 30 predictions and negative rank correlation indicate poor discrimination between acceptable and unacceptable errors.}
\label{fig:platform_estimator}
\end{figure}

The initial error estimator is assessed on the same 30 held-out S3 paths used for the forward-error test. In \cref{fig:platform_estimator}, $\Ehat$ is compared with the FE-evaluated error $\EQ$. The estimator rejects 28 of the 30 predictions. Of the 21 predictions with $\EQ>1$, one is incorrectly accepted. Of the nine predictions with $\EQ\leq1$, eight are incorrectly rejected. The fraction satisfying $\EQ\leq\Ehat$, reported as empirical coverage, is 0.87. The Spearman rank correlation between estimated and actual error is $-0.11$.

The estimator excludes most inaccurate predictions, but it also excludes nearly all accurate predictions in this shifted test. The small number of false acceptances therefore does not establish effective discrimination. When few surrogate predictions are accepted, the hybrid command calculation has little information beyond its four FE evaluations. The estimate remains empirical and does not provide a guaranteed error bound.

\subsubsection{Command selection}

The command tests use the eight fixed target poses and four operating conditions defined in \cref{sec:platform_data,tab:platform_conditions}. The hybrid calculation combines the four selected FE evaluations with accepted surrogate predictions. Its selected command is assessed using the FE loss in \cref{eq:platform_loss} and regret in \cref{eq:regret}.

\begin{figure}[h]
\centering
\includegraphics[width=0.7\linewidth]{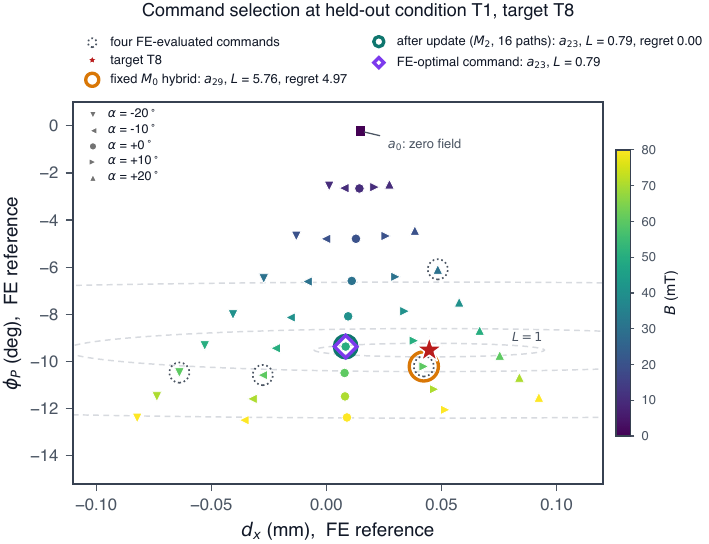}
\caption{FE responses of all 41 platform commands in the $(d_x,\phi_P)$ plane at $T_1$ for the eighth target, labeled T8. The dashed ellipses are contours of the pose loss $L_P$ about T8. The initial hybrid calculation selects command 29 rather than the FE-optimal command 23. The accepted 16-path random update selects command 23 while using the same four FE evaluations. The remaining FE responses are used only to assess the selected commands.}
\label{fig:platform_command}
\end{figure}

\cref{fig:platform_command} illustrates the calculation at $T_1$ for the eighth fixed target, labeled T8 in the figure. The initial-operator calculation selects command 29, whose FE loss is 5.76. The FE-optimal command is 23, with loss 0.79, giving regret 4.97. After the accepted 16-path random update, the hybrid calculation selects command 23 and has zero regret. The same four commands are evaluated by FE before and after updating. The total available command set, including FE-evaluated commands, expands from four to all 41 candidates as more surrogate predictions are accepted.

\begin{figure}[h]
\centering
\includegraphics[width=\linewidth]{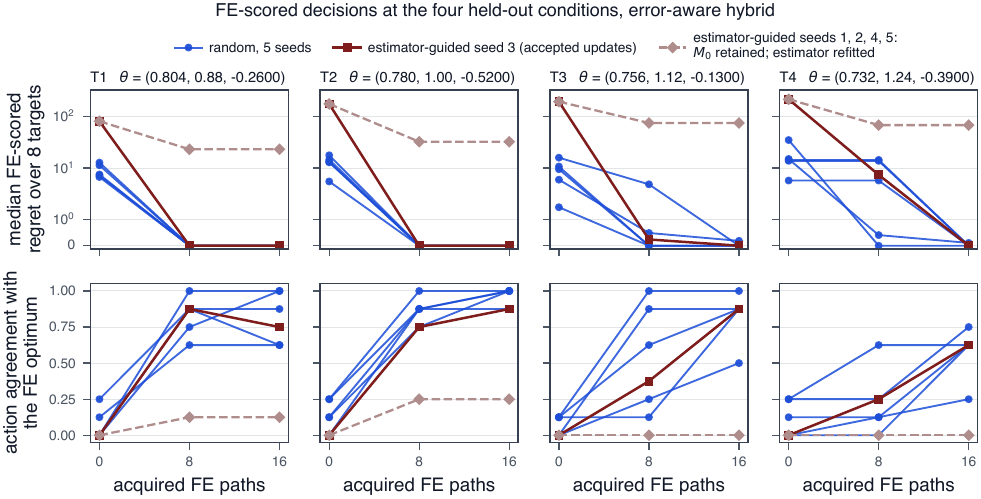}
\caption{Platform command results before adaptation and after eight and 16 acquired paths at the conditions in \cref{tab:platform_conditions}. The upper row shows median FE-evaluated regret over eight targets for each acquisition sequence; the lower row shows the fraction of targets selecting the FE-optimal command. Random and estimator-guided acquisition already select different four-command FE sets before adaptation, so their initial differences are not caused by learning. Sequences whose operator updates fail validation retain $M_0$, although refitting the estimator and selecting a new four-command FE set can change their decisions.}
\label{fig:platform_decisions}
\end{figure}

The complete command results are shown in \cref{fig:platform_decisions}. At each test condition, each acquisition sequence is summarized by its median regret over the eight targets. Taking the median across the five random-acquisition sequences gives 7.51, 13.16, 9.51, and 14.34 before adaptation at $T_1$ through $T_4$, respectively. After eight acquired paths, these values are 0, 0, 0, and 5.74. After 16 paths, the median is zero at every condition. Residual differences remain between sequences: the largest reported seed-level regrets after 16 paths are 0 at $T_1$ and $T_2$, 0.189 at $T_3$, and 0.108 at $T_4$. Zero median regret does not mean that every target selects the FE-optimal command, as the agreement results in the lower row of the figure show.

The accepted estimator-guided sequence, seed 3, reduces its median regret across the four conditions from 182 to 0.121 after eight paths and to zero after 16 paths. The other four estimator-guided sequences retain $M_0$ because their candidate updates fail validation. Changes in their command results arise from refitting the estimator and changing the four commands selected for FE, rather than from updating the forward operator.

\subsubsection{Computational work}

\begin{figure}[h]
\centering
\includegraphics[width=\linewidth]{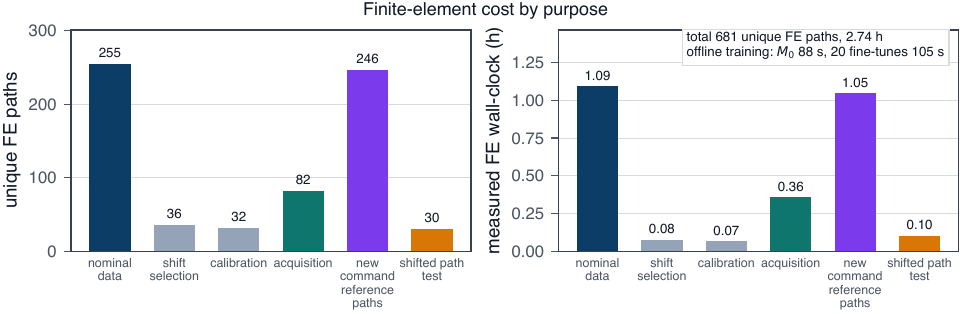}
\caption{Unique platform FE loading paths and recorded FE solution time by purpose. The nominal category includes training, validation, and testing. Complete-command reference calculations are used to assess the selected commands; only four selected FE evaluations enter each hybrid command calculation. These totals do not constitute a measurement of the time required for one complete surrogate-assisted decision.}
\label{fig:platform_cost}
\end{figure}

The platform study uses 681 unique FE loading paths and \SI{2.74}{h} of recorded single-process FE solution time. Their purposes are separated in \cref{fig:platform_cost}. The nominal data comprise 200 training, 25 validation, and 30 test paths, totaling 255. Preliminary evaluation of S1 through S3 uses 36 paths, and initial estimator calibration uses 32. Acquisition contributes 82 unique paths after accounting for reuse, and final shifted forward evaluation uses 30 S3 paths. The complete-command FE response sets contain 328 paths, of which 82 are reused from acquisition and 246 require additional solutions.

Training $M_0$ takes \SI{88}{s}, and the 20 attempts to train updated operators take \SI{105}{s} in total. These measurements quantify FE data generation and operator training. They do not establish an end-to-end speedup because the combined times for neural-operator evaluation, residual assembly, the linearized solve, QoI evaluation, and command selection were not recorded for the final tests.

\section{Gripping wishbone}
\label{sec:wishbone}

The wishbone study applies the same prediction, error-estimation, and adaptation procedure to a different geometry and command objective. The target response is specified by finger aperture and lateral midpoint, while vertical mismatch contributes to the loss and root strain enters the admissibility checks; see \cref{eq:wishbone_qois,eq:wishbone_loss}. Its FE data, POD basis, neural operator, and estimator are constructed independently of those used for the platform.

\subsection{Numerical setup}
\label{sec:wishbone_setup}

\subsubsection{Geometry and FE model}

\begin{figure}[h]
\centering
\includegraphics[width=\linewidth]{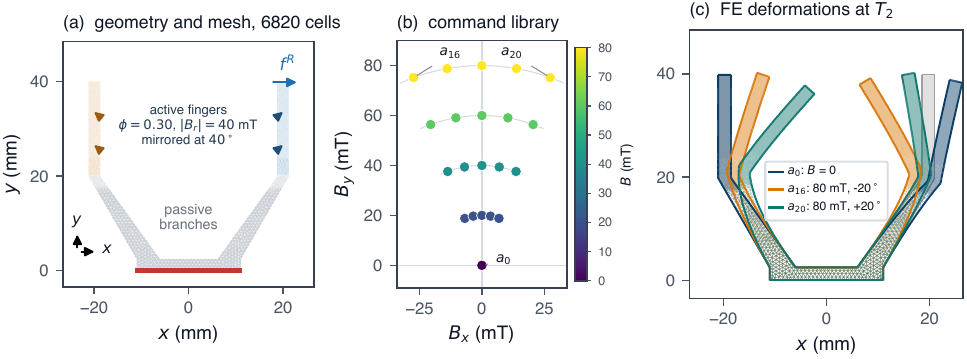}
\caption{Wishbone geometry, candidate commands, and FE responses. (a) Plane-strain geometry and mesh, clamped base, right-tip force, and active and passive regions. The fingers have $\phi=0.30$ and $|\bm B^r|=\SI{40}{mT}$, with mirrored remanent directions indicated by arrows. (b) The 21 candidate commands in the $(B_x,B_y)$ plane. (c) FE deformations at $T_2$ in \cref{tab:wishbone_conditions}: the zero-field command $a_0$ and commands $a_{16}$ and $a_{20}$ with $B=\SI{80}{mT}$ and $\alpha=-20^{\circ}$ and $+20^{\circ}$, respectively. Deformations are shown at true scale over the reference mesh.}
\label{fig:wishbone_setup}
\end{figure}

The wishbone geometry, candidate command set, and representative FE deformations are shown in \cref{fig:wishbone_setup}. The base is clamped, and the prescribed force $(f_x^R,f_y^R)$ is applied to the right tip. The lower branches are passive, with $\phi=0$, while both upper fingers have $\phi=0.30$ and a remanent flux density of magnitude \SI{40}{mT}. The mirrored remanent directions are indicated in the figure. The active fingers meet the passive branches at $y=\SI{20.32}{mm}$ in the reference configuration.

The domain is discretized using 6,820 conforming triangles and vector-valued quadratic Lagrange elements, giving 28,510 displacement degrees of freedom. The equilibrium solution uses Newton iteration with 16 load increments and the MUMPS direct linear solver. Each stored path is checked for nonlinear convergence, satisfaction of the prescribed displacement conditions, finite solution values, positive $J$ at the sampled locations, and absence of finger intersections. Mechanical admissibility additionally requires $g\geq\SI{2}{mm}$ and $q_{99}\leq0.35$. The strain limit is a numerical constraint used in this study, not an experimentally determined failure threshold. The simulations describe free finger motion; contact with a gripped object and the resulting gripping force are not modeled.

\subsubsection{Nominal data, shifted conditions, and targets}
\label{sec:wishbone_data}

Nominal input combinations for the training, validation, and test sets are generated within the following ranges using randomized Sobol sampling, with a different seed for each set:
\begin{equation}
\begin{aligned}
B&\in[0,80]\ \si{mT},\qquad \alpha\in[-20,20]^{\circ},\\
\gamma_G&\in[0.90,1.10],\qquad f_x^R,f_y^R\in[-0.75,0.75]\ \si{mN}.
\end{aligned}
\label{eq:wishbone_nominal_box}
\end{equation}
The 200 training, 25 validation, and 30 test paths are disjoint. The shifted operating range is
\begin{equation}
\gamma_G\in[0.75,0.89],\qquad f_x^R\in[1.00,2.00]\ \si{mN},\qquad f_y^R\in[0.75,1.50]\ \si{mN},
\label{eq:wishbone_shifted_box}
\end{equation}
with the magnetic-field ranges unchanged. Thirty held-out paths are generated from this shifted range using scrambled Sobol seed 26092101. These paths are excluded from estimator fitting, FE acquisition, continued training, and selection of the updated operators or training stopping criteria.

For the wishbone, \cref{tab:wishbone_conditions} defines four adaptation conditions, $A_1$ through $A_4$, and four held-out command-test conditions, $T_1$ through $T_4$. Four FE paths are acquired at each $A_k$. The test conditions lie at increasing normalized distances $d_{\mathrm{box}}$ from the nominal input range, where the distance is defined in \cref{eq:box_distance}. The paths acquired at the adaptation conditions provide adaptation data, while FE solutions at the test conditions provide separate reference responses for final evaluation.

\begin{table}[h]
\centering
\caption{Wishbone operating conditions. Force components are in mN. Four FE paths are acquired at each $A_k$; the conditions $T_1$ through $T_4$ are reserved for testing command selection. The last column gives the normalized distance outside the nominal input box.}
\label{tab:wishbone_conditions}
\begin{tabular}{@{}llrrrr@{}}
\toprule
Condition & Role & $\gamma_G$ & $f_x^R$ & $f_y^R$ & $d_{\mathrm{box}}$\\
\midrule
$A_1$ & Adaptation & 0.85 & 1.25 & 0.80 & 0.4180\\
$A_2$ & Adaptation & 0.83 & 1.40 & 0.90 & 0.5659\\
$A_3$ & Adaptation & 0.80 & 1.75 & 1.25 & 0.8975\\
$A_4$ & Adaptation & 0.78 & 2.00 & 1.50 & 1.1421\\
\midrule
$T_1$ & Command test & 0.8693 & 1.0011 & 1.1006 & 0.3259\\
$T_2$ & Command test & 0.8086 & 1.1120 & 1.1309 & 0.5757\\
$T_3$ & Command test & 0.7685 & 1.5324 & 0.9138 & 0.8462\\
$T_4$ & Command test & 0.7543 & 1.6733 & 1.4901 & 1.0739\\
\bottomrule
\end{tabular}
\end{table}

The condition $T_1$ is separated by a normalized input distance of 0.066 from an earlier development condition used to select the estimator procedure. Its FE responses remain held out, but this proximity limits the conclusions that can be drawn about estimator performance away from the development conditions. The other three test conditions are at least 0.253 from the nearest earlier adaptation or decision condition.

The eight target apertures and lateral midpoints are generated at $\bm\theta_{\mathrm{ref}}=(1,0,0)$ using magnetic commands outside the 21-command candidate set, as described in \cref{sec:problem}. These targets remain fixed across $T_1$ through $T_4$. Evaluating all 21 commands at the four test conditions gives 84 FE loading-path endpoints. At each condition, the same response set is used for all eight targets, giving 32 command-selection problems per operator. Separate command tests at four nominal operating conditions also use the eight targets to assess performance before the operating inputs are shifted.

\subsubsection{Training, estimator data, and evaluation}
\label{sec:wishbone_evaluation}

The initial operator $M_0$ uses the rank-16 representation and training settings in \cref{sec:method,tab:initial_training}. Updates after eight and 16 acquired paths give $M_1$ and $M_2$. Five training seeds are used for each acquisition strategy. Estimator-guided acquisition selects the same four command indices, $(0,20,16,18)$, at every adaptation condition, producing one distinct set of 16 paths. The five random-acquisition sequences produce five distinct path sets. Thus, estimator-guided acquisition is assessed through repeated training on one acquired dataset, while the random-acquisition comparison also varies the acquired data.

Separate error estimators are fitted for acquisition and final assessment. For each operator, the estimator used for final assessment is fitted using 146 development endpoints: 32 initial calibration endpoints, 30 earlier shifted endpoints, and 84 endpoints from complete 21-command evaluations at four earlier adaptation conditions. These data are divided into regression and quantile-calibration subsets of 73 endpoints each. The division is made by complete path or operating condition rather than by individual samples within those groups.

At condition $A_k$, the acquisition estimator uses only FE data available before the four new commands are selected. It cannot use the complete FE response set at that condition to choose which commands to acquire. Neither estimator fit uses the 30 final shifted endpoints, the 84 final command-test endpoints, or the 30 nominal test endpoints. The estimator procedure and the requirements for accepting an updated operator are fixed using development data before final testing.

Forward accuracy and estimator reliability are assessed on the nominal and shifted path tests and on the 21 candidate commands at each $T_k$. Command agreement and regret are evaluated over the eight targets at each condition. Results are calculated separately for each trained operator before taking a median across the five training seeds. The fixed operator and the five $M_1$ and five $M_2$ variants for each of the two acquisition strategies give 21 operator variants, or 84 combinations of operator and operating condition over the four shifted test conditions. 

For the field comparisons, each 30-path test set is ordered by increasing $\EQ$ for the initial operator, and the 16th path is displayed. This is the upper of the two middle paths. The same shifted input is retained when comparing the initial and updated operators, with training seed 1 used for the illustrated update. The remaining results use all five seeds.

\FloatBarrier
\subsection{Results}
\label{sec:wishbone_results}

\subsubsection{Nominal forward accuracy}

Within the nominal range in \cref{eq:wishbone_nominal_box}, the initial wishbone operator has median $\EQ=0.140$ on the 30 held-out test paths, with 29 predictions satisfying $\EQ\leq1$. In the separate command tests at four nominal operating conditions described in \cref{sec:wishbone_data}, it selects the FE-optimal command for all eight targets and has zero mean regret.

\begin{figure}[h]
\centering
\includegraphics[width=\linewidth]{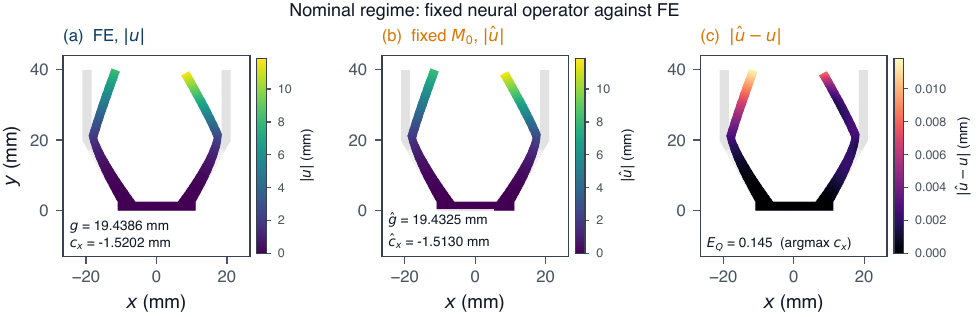}
\caption{Wishbone FE solution and initial prediction for the 16th of 30 nominal test paths ordered by the initial operator's $\EQ$. Panels (a) and (b) use a common displacement-magnitude scale; panel (c) shows the pointwise displacement error on a separate scale. Fields are shown on their deformed configurations at true scale over the reference mesh.}
\label{fig:wishbone_nominal_field}
\end{figure}

\cref{fig:wishbone_nominal_field} shows the 16th path after ordering the nominal test set by $\EQ$, following \cref{sec:wishbone_evaluation}. Its normalized QoI error is 0.145. The FE and predicted deformation fields are visually similar at the plotted scale, and their pointwise difference is much smaller than the displacement magnitude.

\subsubsection{Shifted forward accuracy and adaptation}

Under the reduced stiffness and increased tip forces in \cref{eq:wishbone_shifted_box}, median $\EQ$ on the 30 held-out paths increases from 0.140 to 5.097, and only three predictions satisfy the tolerance. The separate command tests show increasing error along $T_1$ through $T_4$ in \cref{tab:wishbone_conditions}. The medians over the 21 candidate commands are 1.353, 2.299, 5.702, and 11.373, with 9, 3, 0, and 0 predictions satisfying $\EQ\leq1$, respectively.

\begin{figure}[h]
\centering
\includegraphics[width=\linewidth]{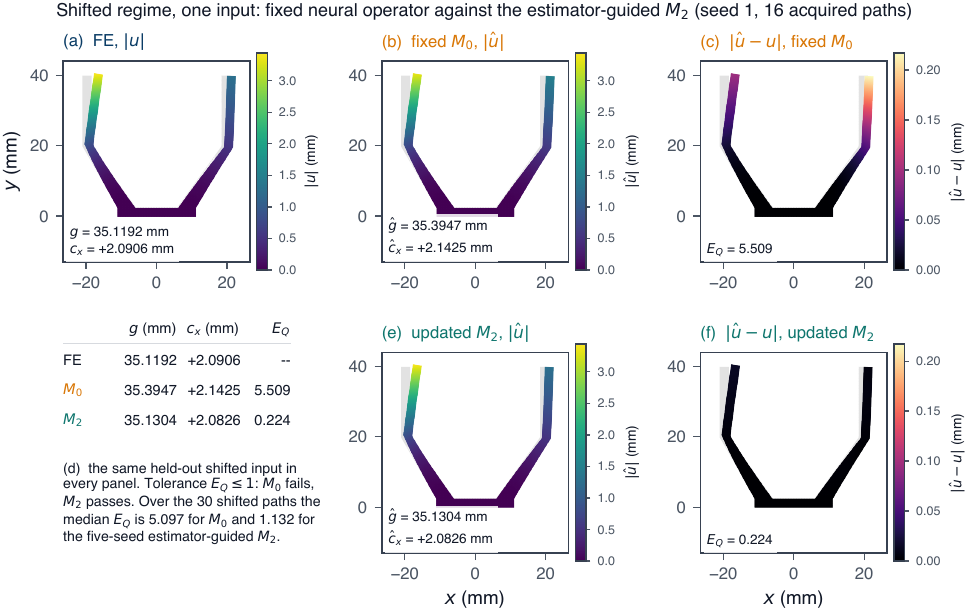}
\caption{Wishbone FE solution, initial prediction, and 16-path updated prediction for the same held-out shifted input. The input is the 16th of 30 paths ordered by the initial operator's $\EQ$, and the illustrated update uses estimator-guided training seed 1. Panels (a), (b), and (e) share one displacement scale; panels (c) and (f) share one error scale. Deformations are shown at true scale. Results across all five seeds are reported in \cref{tab:wishbone_forward}.}
\label{fig:wishbone_ood_field}
\end{figure}

\cref{fig:wishbone_ood_field} compares FE, initial-operator, and updated-operator responses at the 16th input after ordering the 30 shifted test paths by the initial error. The initial prediction gives $\EQ=5.51$, with visible differences in both deformation and displacement magnitude. For the same input, the estimator-guided $M_2$ operator from training seed 1 gives $\EQ=0.22$. This is one illustrative comparison rather than the aggregate error after adaptation.

\cref{tab:wishbone_forward} separates the 30-path shifted test from the four command-test conditions. For each updated operator, the median is first calculated over the relevant paths or commands, followed by the median across the five training seeds. On the shifted path test, estimator-guided acquisition reduces this value from 5.097 to 3.439 after eight paths and to 1.132 after 16 paths. The corresponding 16-path value for random acquisition is 0.803. The second estimator-guided update does not improve every test condition: the median at $T_1$ increases from 0.785 to 1.024, while those at $T_2$, $T_3$, and $T_4$ decrease.

\begin{table}[h]
\centering
\caption{Wishbone normalized QoI errors before and after adaptation. The shifted-path column uses the 30 test paths from \cref{eq:wishbone_shifted_box}; each $T_k$ column uses the 21 candidate commands at that condition in \cref{tab:wishbone_conditions}. Entries for $M_0$ are medians over the corresponding inputs. Updated-operator entries are medians across five training seeds of those input-wise medians.}
\label{tab:wishbone_forward}
\small
\begin{tabular}{@{}lrrrrr@{}}
\toprule
Operator & Shifted paths & $T_1$ & $T_2$ & $T_3$ & $T_4$\\
\midrule
$M_0$ & 5.097 & 1.353 & 2.299 & 5.702 & 11.373\\
$M_1$, estimator-guided & 3.439 & 0.785 & 1.192 & 3.030 & 7.532\\
$M_2$, estimator-guided & 1.132 & 1.024 & 0.770 & 0.775 & 1.131\\
$M_2$, random & 0.803 & 0.946 & 0.819 & 0.878 & 1.149\\
\bottomrule
\end{tabular}
\end{table}

The updated operators retain accuracy on the separate nominal test. The five estimator-guided $M_2$ medians range from 0.134 to 0.206, with all 30 predictions satisfying the tolerance for every operator. The five random-acquisition $M_2$ medians range from 0.101 to 0.175; four operators satisfy the tolerance on all 30 paths and one on 29 paths. Random acquisition gives a smaller aggregate shifted error in this experiment, but the comparison uses one estimator-guided path set and five random path sets, as explained in \cref{sec:wishbone_evaluation}. It does not establish a general advantage of either acquisition strategy.

\subsubsection{QoI-error estimation}

The estimator associated with $M_0$ has no false acceptance on the nominal test paths, the shifted test paths, or the candidate commands at the four test conditions in \cref{tab:wishbone_conditions}. Its Spearman rank correlation with $\EQ$ is 0.881 on the shifted paths and 0.730, 0.761, 0.816, and 0.748 at $T_1$ through $T_4$, respectively. It nevertheless accepts almost no surrogate prediction at the shifted command-test conditions, leaving command selection largely to the four FE evaluations.

After adaptation, the true error and fitted estimate both change. For estimator-guided $M_2$, no false acceptance occurs at $T_2$ or $T_3$ in any of the five training seeds. At $T_1$, every seed falsely accepts between one and three predictions. At $T_4$, only one seed accepts a surrogate prediction beyond the four FE-evaluated commands; one of its two accepted predictions is inaccurate. The relatively small distance between $T_1$ and an earlier estimator-development condition, noted in \cref{sec:wishbone_data}, further limits its use as evidence of estimator reliability under unfamiliar conditions.

The estimator can also reject accurate predictions after adaptation. One updated operator has 30 accurate nominal test predictions, but its estimator accepts only 12. This shows that improved forward accuracy does not consistently lead to reliable prediction acceptance.

\subsubsection{Command selection}

The effect of estimator acceptance on the selected commands is assessed by comparing the hybrid and four-FE calculations defined in \cref{sec:method}. Both use the same four FE evaluations at a given test condition; the hybrid calculation additionally uses accepted surrogate predictions. For each trained operator and condition, agreement is the number of the eight targets for which the FE-optimal command is selected, and mean regret is calculated over the same eight targets. \cref{tab:wishbone_commands} reports the median of each quantity across the five estimator-guided $M_2$ training seeds.

At $T_2$, four of the five seeds accept at least one surrogate prediction beyond the FE-evaluated commands. The hybrid calculation increases median agreement from one to three targets and reduces the reported mean regret from 1.213 to 0.503. At $T_3$, three seeds use additional surrogate predictions; median agreement increases from two to three targets, and mean regret decreases from 1.174 to 0.360. Neither condition has a false acceptance in any of the five seeds. These intermediate shifts therefore show an improvement over selecting from the four FE-evaluated commands alone.

\begin{table}[h]
\centering
\caption{Wishbone command selection using estimator-guided $M_2$ operators at the conditions in \cref{tab:wishbone_conditions}. Agreement counts FE-optimal selections among eight targets. Regret is averaged over eight targets for each operator. Both statistics are then summarized by the median across five training seeds. The last column gives the number of seeds using additional accepted surrogate predictions and indicates whether false acceptance occurs.}
\label{tab:wishbone_commands}
\small
\begin{tabular}{@{}lrrrrl@{}}
\toprule
& \multicolumn{2}{c}{Agreement} & \multicolumn{2}{c}{Mean regret} & \\
\cmidrule(lr){2-3}\cmidrule(lr){4-5}
Condition & Hybrid & Four FE & Hybrid & Four FE & Surrogate use\\
\midrule
$T_1$ & 5 & 1 & 0.045 & 1.099 & 5/5; false acceptance\\
$T_2$ & 3 & 1 & 0.503 & 1.213 & 4/5; no false acceptance\\
$T_3$ & 3 & 2 & 0.360 & 1.174 & 3/5; no false acceptance\\
$T_4$ & 2 & 2 & 1.031 & 1.031 & 1/5; one false acceptance\\
\bottomrule
\end{tabular}
\end{table}

The largest reduction in reported mean regret occurs at $T_1$, but the false-acceptance counts are 2, 3, 1, 3, and 3 across the five seeds. The lower loss does not therefore establish reliable acceptance of the surrogate predictions. At the most severe test condition, $T_4$, the median hybrid result equals the four-FE result. Thus, inaccurate predictions can be accepted at the mildest shift, while insufficient surrogate acceptance limits the benefit at the most severe shift. Across all 84 combinations of operator and operating condition defined in \cref{sec:wishbone_evaluation}, every selected command is FE-admissible and every target has at least one available candidate.

\begin{figure}[h]
\centering
\includegraphics[width=\linewidth]{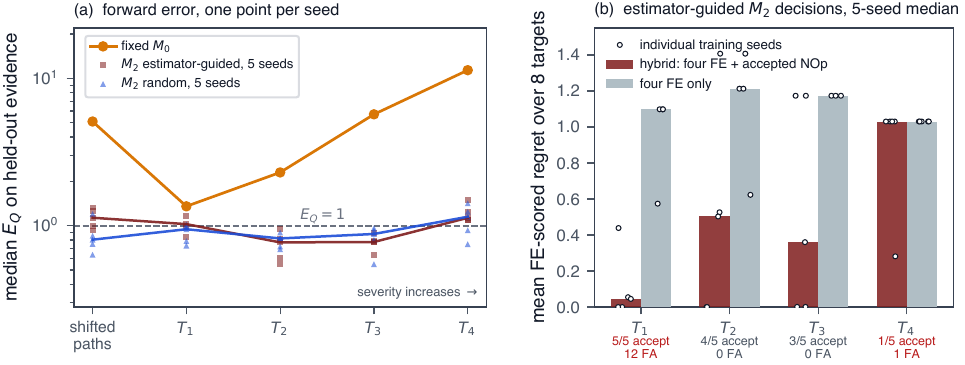}
\caption{Wishbone forward accuracy and command selection after adaptation. (a) Median $\EQ$ on the shifted path test and the four conditions in \cref{tab:wishbone_conditions}. For the updated operators, individual markers show training seeds and the lines give five-seed medians; the $M_0$ curve represents the single fixed operator. (b) Mean regret over eight targets for the estimator-guided $M_2$ hybrid and four-FE calculations; open circles show individual seeds, and bars give the median across seeds. Annotations report the number of seeds using additional surrogate predictions and total false acceptances.}
\label{fig:wishbone_summary}
\end{figure}

\subsubsection{Computational work}

The final shifted evaluation uses 114 new FE paths: 30 for the forward test and 84 for the complete candidate command sets at the four test conditions. All satisfy the prescribed FE checks, and their recorded solution time is \SI{652}{s}. For training seed 1, an estimator-guided $M_2$ update using stored paths takes \SI{12.5}{s}. These measurements distinguish the time required for training an updated operator from that required to generate reference FE data. They do not establish an end-to-end speedup because the combined times for residual assembly, the linearized solve, QoI evaluation, and command selection were not measured under a common timing procedure.

\section{Discussion}
\label{sec:discussion}

The two studies support the reuse of FE loading paths to adapt neural operators under the prescribed changes in stiffness and mechanical loading. Although the structures have different geometries, QoIs, and material parameters, adaptation improves forward accuracy while preserving accuracy in the nominal regime. The effects on error estimation and command selection are less consistent. \cref{tab:cross_example_comparison} summarizes these distinctions.

\begin{table}[htbp]
\centering
\small
\caption{Comparison of the principal findings. Updated forward errors correspond to 16 randomly acquired paths and are summarized by the median of the five per-operator median errors. The examples use different QoIs, tolerances, and shifted operating ranges.}
\label{tab:cross_example_comparison}
\renewcommand{\arraystretch}{1.10}
\begin{tabularx}{\linewidth}{@{}>{\raggedright\arraybackslash}p{0.20\linewidth}>{\raggedright\arraybackslash}X>{\raggedright\arraybackslash}X@{}}
\toprule
\textbf{Aspect} & \textbf{Platform} & \textbf{Wishbone}\\
\midrule
Forward adaptation & Shifted median $\EQ$ decreases from 1.147 to 0.144 & Shifted median $\EQ$ decreases from 5.097 to 0.803\\
\addlinespace
Error estimation & Initial estimator rejects most shifted predictions but poorly distinguishes their errors & Some estimators falsely accept predictions after adaptation\\
\addlinespace
Command selection & Zero median regret at all four test conditions after adaptation; individual errors remain & Hybrid improves on the four-FE calculation at $T_2$ and $T_3$; equal median regret at $T_4$\\
\addlinespace
Distinct acquired data & Two estimator-guided and five random 16-path sets & One estimator-guided and five random 16-path sets\\
\bottomrule
\end{tabularx}
\end{table}

The platform projection test identifies why adaptation within the fixed POD space is effective for the tested inputs. Under the S3 shift in \cref{tab:platform_shift}, the reduced basis represents the FE solutions with substantially smaller QoI error than the initial neural operator. The dominant error therefore arises from the learned input-to-coefficient map rather than basis truncation. Updating this map improves predictions without modifying the reduced representation. This improvement is distinct from residual correction \cite{cao2023residual,jha2024residual}: the linearized residual solve in \cref{eq:linearized_indicator} supplies estimator information but does not modify the displacement used for command selection.

Forward accuracy, estimator reliability, and command quality must nevertheless be assessed separately. The initial platform estimator excludes most inaccurate predictions but also rejects nearly all accurate shifted predictions. For the wishbone, the adapted operators improve command selection at $T_2$ and $T_3$ (\cref{tab:wishbone_conditions}) without false acceptance in the tested cases. At $T_1$, however, lower regret is accompanied by false acceptance, while at $T_4$ the median hybrid regret equals that of the four-FE calculation. Thus, reduced command loss does not by itself establish reliable prediction acceptance, and improved forward accuracy does not ensure a benefit at every operating condition.

The acquisition comparison is limited by the number of distinct data sets in \cref{tab:cross_example_comparison}. Repeated training on the same acquired paths evaluates training variability, not additional acquisition outcomes. For the platform, the validation requirements also determine which candidate updates are used in subsequent decisions. The results therefore support adaptation using acquired FE paths but do not establish an advantage of estimator-guided acquisition over random sampling.

These conclusions apply to two-dimensional hMSM models with fixed rank-16 representations, prescribed input ranges, discrete command sets, and 30-path forward tests. The physical systems are simulated by FE with known inputs, so experimental uncertainty and sensor-based inference are outside the study. The demonstrated benefit is improved prediction and command quality under a fixed FE budget; reliable acceptance of the adapted predictions remains a separate requirement.

\section{Conclusions}
\label{sec:conclusion}

This work studied periodic adaptation of neural operators for command selection in two hard-magnetic soft-material structures under changing known operating conditions. The physical systems were simulated using high-fidelity FE models. Four selected FE evaluations per operating condition supplemented the surrogate predictions, while their complete loading paths were retained for updating the neural operator and its empirical QoI-error estimator.

Both initial operators lost substantial accuracy when stiffness and mechanical loading moved beyond their training ranges. Updates using 16 acquired FE paths recovered much of this accuracy while preserving accuracy in the nominal regime. The improved predictions also supported better command selection under the same FE budget, although the benefit depended on the operating condition. The acquisition comparisons did not establish an advantage of estimator-guided selection over random selection.

The empirical estimators remained unreliable at some operating conditions after adaptation, accepting inaccurate predictions or rejecting accurate ones. Improved forward accuracy therefore does not by itself establish which predictions are suitable for command selection. The study demonstrates that FE solutions generated during repeated mechanics calculations can improve subsequent predictions and decisions through data reuse. Reliable use of these adapted predictions additionally requires error assessment that remains effective as both the operating conditions and the learned approximation change.

\section*{Acknowledgments}
PKJ acknowledges support from the National Science Foundation through the Engineering Research Initiation (ERI) program under award No.~2502279, and from the South Dakota Board of Regents Competitive Research Grant (SDBOR CRG) program. These awards supported IG's and HA's work on this project at different stages. Any opinions, findings, and conclusions or recommendations expressed in this material are those of the authors and do not necessarily reflect the views of the funding agencies.

\section*{Data and code availability}
The source and supporting data will be released with the accepted manuscript.

\section*{Declaration of competing interest}
The authors declare no competing interests.

\bibliographystyle{unsrt}
\bibliography{references}

\end{document}